\documentclass[twocolumn,aps,english,prb,showpacs,floatfix]{revtex4-1}
\usepackage[colorlinks=true,urlcolor=blue,citecolor=blue,linkcolor=blue]{hyperref}
\usepackage[sort&compress]{natbib}
\usepackage[T1]{fontenc}
\usepackage[latin9]{inputenc}
\usepackage{amssymb}
\usepackage{graphicx}
\usepackage{amsmath,color}
\usepackage{mathrsfs}
\usepackage{float}
\usepackage{indentfirst}
\usepackage{babel}
\usepackage{color}

\usepackage[sort&compress]{natbib}
\setcitestyle{numbers,square}

\makeatletter

\begin{document}

\title{Optimizing spin-orbit splittings in InSb Majorana nanowires}
\author{Alexey A. Soluyanov}
\author{Dominik Gresch}
\affiliation{Theoretical Physics and Station Q Zurich, ETH Zurich, 8093 Zurich, Switzerland}
\author{Roman M. Lutchyn}
\author{Bela Bauer}
\author{Chetan Nayak}
\affiliation{Station Q, Microsoft Research, Santa Barbara, California 93106, USA}
\author{Matthias Troyer}
\affiliation{Theoretical Physics and Station Q Zurich, ETH Zurich, 8093 Zurich, Switzerland}

\date{\today}

\begin{abstract}
Semiconductor-superconductor heterostructures represent a promising platform for the detection of Majorana zero modes and subsequently the processing of quantum information using their exotic non-Abelian statistics. Theoretical modeling of such low-dimensional heterostructures is generally based on phenomenological effective models. However, a more microscopic understanding of the band structure and, especially, of the spin-orbit coupling of electrons in these devices is important for optimizing their parameters for applications in quantum computing. In this paper, we approach this problem by first obtaining a highly accurate effective tight-binding model of bulk InSb from {\it ab initio} calculations. This model is symmetrized and correctly reproduces both the band structure and the wavefunction character. It is then used to simulate slabs of InSb in external electric fields. The results of this simulation are used to determine a growth direction for InSb nanowires that optimizes the conditions for the experimental realization of Majorana zero modes.    
\end{abstract}
\maketitle

\section{Introduction}
The search for Majorana zero modes (MZMs) in solid-state systems attracted a lot of interest~\cite{Reich, Brouwer_Science, Wilczek2012, LeeYazdani} due to the theoretical prediction that defects binding MZMs manifest non-Abelian quantum statistics~\cite{Moore1991,Nayak1996, ReadGreen, Ivanov}, and, as such, would open up the possibility to realize topological quantum computing~\cite{Kitaev-Ann03, Nayak-RMP08, Alicea12, BeenakkerReviewA, NayakReview2015}. Topological quantum computing schemes use topological degrees of freedom to encode information. Since topological degrees of freedom are inherently non-local, they do not couple to local operations. Therefore, the error rates in topological quantum computing schemes are exponentially suppressed with distance between anyons (i.e. MZMs) and inverse temperature, providing an enormous advantage over conventional quantum computing platforms.

In condensed matter physics, MZMs were first discussed in the context of fractional quantum Hall effect~\cite{Moore-NPB91} and topological $p$-wave superconductors/superfluids~\cite{ReadGreen, Volovik99, Kitaev-Usp01, Volovik-book, Volovik-JETPL09}. Later on, it was shown that topological superconductivity can be realized in various heterostructures~\cite{Fu-PRL08, Sau, Alicea, Lutchyn-PRL10, Oreg-PRL10, Mourik2012, Nadj-Science14}. In particular, semiconductor-superconductor heterostructures are very promising, and arguably the simplest, experimental systems for realizing MZMs. Indeed, the recipe for engineering topological superconductivity in semiconductor-superconductor heterostructures involves three main ingredients: spin-orbit coupling (SOC), Zeeman splitting, and proximity induced $s$-wave pairing~\cite{Sau, Alicea, Lutchyn-PRL10, Oreg-PRL10}. The appropriate combination of these ingredients leads to an effective Hamiltonian equivalent to that of a spinless $p$-wave superconductor.

Theoretical predictions for realizing topological superconductivity in semiconductor nanowires coupled to conventional s-wave superconductors~\cite{Lutchyn-PRL10, Oreg-PRL10} have sparked significant experimental activity~\cite{Mourik2012, Rokhinson2012, Das2012, Deng2012, Fink2012, Churchill2013, Chang_PRL2013, Lee_arxiv2013, Deng-ScRep14}.  The first tunneling spectroscopy experiment aiming to detect MZMs was performed in Delft with InSb zincblende nanowires that were proximity-coupled to NbTiN~\cite{Mourik2012}. Later on, the observation of zero-bias peaks in finite magnetic field consistent with the theoretical predictions~\cite{ZeroBiasAnomaly0,ZeroBiasAnomaly1,ZeroBiasAnomaly2,ZeroBiasAnomaly3,ZeroBiasAnomaly31, ZeroBiasAnomaly4,ZeroBiasAnomaly5,ZeroBiasAnomaly6, 1DwiresLutchyn2, ZeroBiasAnomaly61, ZeroBiasAnomaly7} was reported by many other experimental groups~\cite{Das2012, Deng2012, Fink2012, Churchill2013}. Device fabrication process used in Refs.~\cite{Mourik2012, Das2012, Deng2012, Fink2012, Churchill2013} involved self-assembled nanowires as the basis for the semiconducting part, contacted with an s-wave superconductor. It was later found that a better approach is to use molecular beam epitaxy to grow core-shell nanowires with a semiconducting core and a metallic shell, which was successfully implemented in InAs-Al heterostructures~\cite{Krogstrup-NatMat15, Chang2015-NatNano15}.

\begin{figure}[b]
\begin{center}
\includegraphics[width=\columnwidth]{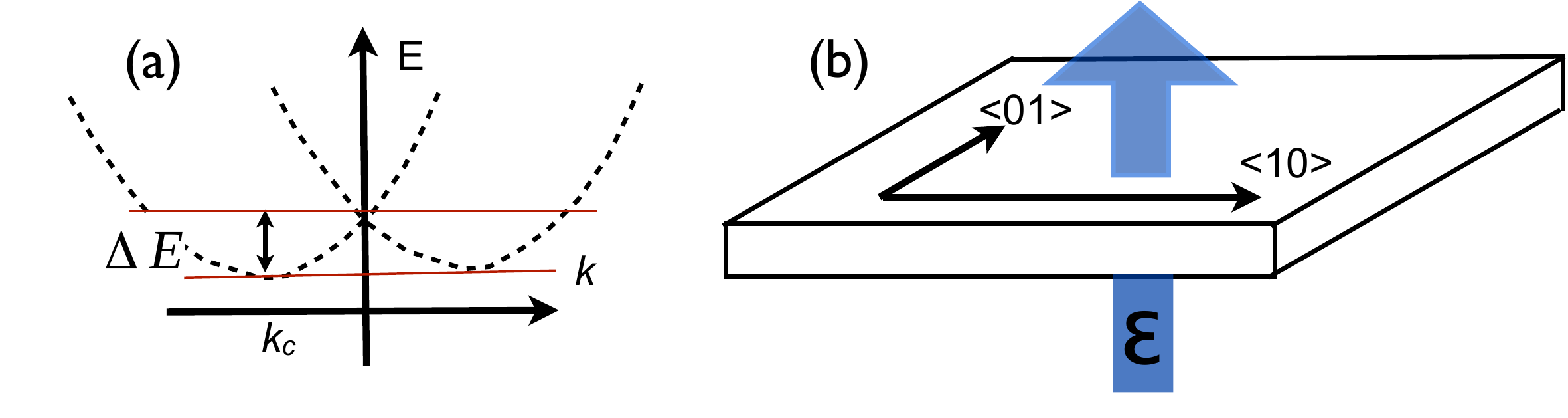}
\end{center}
\caption{(a) Spin splitting $\Delta E$. (b) A slab of InSb put in external electric field, orthogonal to the slab. }
\label{fig:slab}
\end{figure}

Realization of MZMs in both contact and epitaxially proximitized semiconducting wires depends crucially on the strength of SOC~\cite{Lutchyn-PRL10, Oreg-PRL10}. In particular, of major importance for zincblende (ZB) and wurtzite wires in current experimental setups is the size of the spin splitting $\Delta E$ in the first conduction band near the band minimum at the Brillouin zone (BZ) center (see Fig.~\ref{fig:slab}(a)).
It is thus desirable to optimize the heterostructures and predict the growth conditions for which this spin splitting is maximized.

In this paper we address this optimization problem using the example of ZB InSb as a prototypical material for MZM realization. Although the ZB structure lacks inversion symmetry, the spin splitting of the first conduction band in the bulk material is only cubic in momentum ${\bf k}$, resulting in negligible $\Delta E$ in the bulk~\cite{Dresselhaus-PhysRev55}. Thus, it is necessary to generate linear in momentum spin splitting terms in the Hamiltonian. Such terms arise in low-dimensional structures~\cite{Winkler-book} due to bulk inversion asymmetry (BIA) that gives linear in ${\bf k}$ contributions to spin splittings as a result of quantum confinement, and structural inversion asymmetry (SIA) that is associated with the effective electric field that arises due to the macroscopic asymmetry of the structure~\cite{Bychkov-JETPL84}. This effective field can be due to the changes in the crystal potential, or a real macroscopic electric field, applied, for example by gating.

In stand-alone slabs or quantum wells of InSb both BIA and SIA are generally present. The two contributions can combine constructively or destructively depending on the direction of the slab~\cite{Nestoklon-PRB08, Nestoklon-PRB12}, as was shown in photogalvanic experiments for $(001)$ (that is, orthogonal to the $[001]$ direction) ZB GaAs quantum wells~\cite{Ganichev-PRL04, Ganichev-PSS14}. In wires the BIA and SIA terms, being hard to distinguish, combine into a single effective term. It is convenient to think of the wire as cut out of a slab. In this case the full effective linear in ${\bf k}$ splitting becomes a combination of the term that is present in the slab and an additional term arising from confinement of a 2D system to 1D.

Various studies of InSb band structure were carried out since the late fifties~\cite{Kane-JPhCh57}. Most of these studies were limited to symmetry-dictated ${\bf k}\cdot{\bf p}$ expansions~\cite{Kane-kp66} around the center of the BZ~\cite{Cardona-JPhCh63, Cardona-PRB88}. These models are tailored to match the experimentally known features of the band structure exactly, however they are limited to a close vicinity of a special point in ${\bf k}$-space. The next step of approximation is provided by the tight-binding (TB) approximation, which is capable of describing selected bands throughout the BZ. A set of models with short range hoppings, called empirical TB (ETB) models, which are based on symmetry alone, with the hopping parameters matched to reproduce experimental data, is often used in the field of semiconductors~\cite{Slater-PR54, Chadi-PSS75, Vogl-JPCS83, Jancu-PRB98, Klimeck-Superlat00, Luisier-PRB06}. These models have a drawback of not reproducing the wavefunction correctly.

Since the correct representation of the electronic wavefunction is required when aiming to simulate MZMs in a realistic setup, the TB model in the current work is derived from an {\it ab initio} calculation. This model, having long-range hoppings, accurately reproduces the energy spectrum of InSb and by derivation is matched to the wavefunction obtained in the {\it ab initio} calculation. To achieve the required accuracy to reliably extract subtle effects such as spin splittings, our {\it ab initio} simulations are performed using a modification of the HSE06 hybrid scheme~\cite{Heyd-JChPh03,Heyd-JChPh04,Heyd-JChPh06}, and additional symmetrization of the resultant TB model is done, allowing for a band structure description on a sub-meV scale. This method of constructing TB models can be applied generally even for more complicated materials, for which ETBs are not readily available, and can be used to compute SOC effects, as illustrated below. It can also be used for a high-throughput search of materials that can be more suitable for realizing MZMs.

To optimize spin splittings, the obtained TB model is used to simulate different stand-alone slabs of InSb. We find that in the absence of surface relaxation confinement alone results in spin splittings that vanish quickly with increasing slab size. External electric field is further applied orthogonal to the slab (see Fig.~\ref{fig:slab}(b)), and the splitting $\Delta E$ is studied in various directions in 2D momentum space. Arguments are presented that allow us to draw conclusions about optimal directions for the wire growth to maximize $\Delta E$. While the actual numbers provided for the splittings will change depending on the experimental setup, the dependence of susceptibility of spin-splittings to the external fields on the wire growth direction is expected to be a universal property~\footnote{Disorder effects can lead to additional coordinate dependence of spin-splitting terms in the SOC Hamiltonian~\cite{Glazov-PhysE10, Glazov-PRL11}, but we do not consider these effects in the present work.}. 

The paper is organized as follows: In Sec.~\ref{sec:fp} we present our first-principles simulations, followed in Sec.~\ref{sec:tb} by the derivation of TB models. In Sec.~\ref{sec:so}, we present results obtained for spin splittings in finite systems. Conclusions are presented in Sec.~\ref{sec:end}.

\section{First-principles calculations}
\label{sec:fp}

\begin{figure}
\begin{center}
\includegraphics[width=0.6\columnwidth]{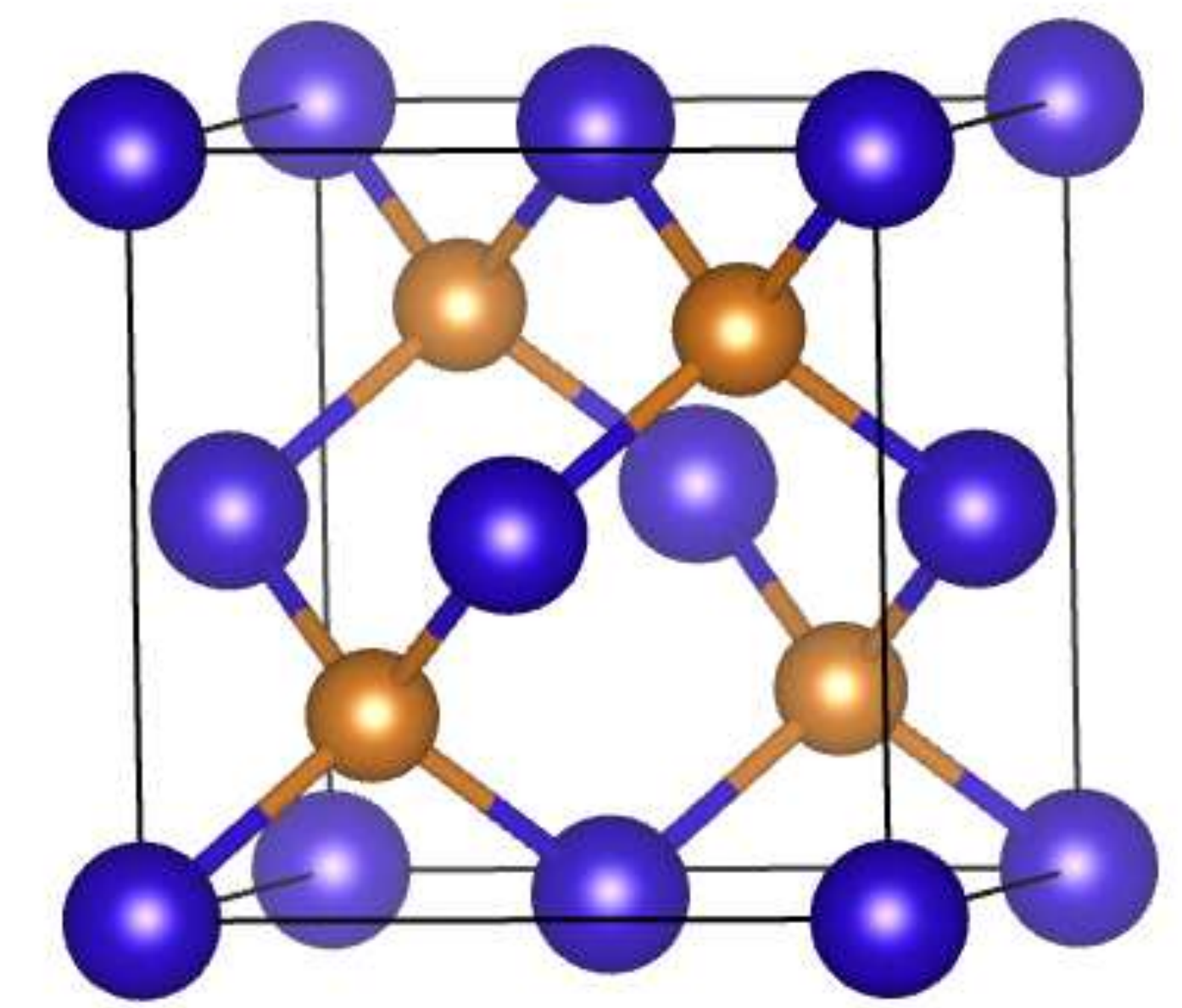}
\end{center}
\caption{Crystal structure of InSb. The conventional FCC unit cell is shown with In in blue and Sb in brown.}
\label{fig:cryst}
\end{figure}

The crystal structure of InSb is ZB, which is equivalent to face-centered cubic (FCC) with one formula unit per primitive unit cell. The conventional unit cell consists of four primitive unit cells as shown in Fig.~\ref{fig:cryst}. The space group is $T_d$ ($\#$216).

To obtain the TB model of InSb, we carried out first-principles calculations using hybrid functionals. It is well known that simpler approximations, like the local density approximation (LDA) or the generalized gradient approximation (GGA), lead to a metallic band structure that results from incorrect band ordering at the $\Gamma$-point~\cite{Massidda-PRB90}. Hybrid functionals~\cite{Muscat-ChPhLett01} and $GW$~\cite{Hedin-PR65,Hedin-SSP70} can be used to fix the band ordering~\cite{Kim-PRB09, Zhu-PRB91,Chantis-PRL06}, and both schemes are generally accepted to be very reliable for computing band gaps of semiconductors nowadays.

In particular, the HSE03/HSE06 hybrid functionals~\cite{Heyd-JChPh03,Heyd-JChPh04,Heyd-JChPh06} proved to be successful in computing band structures of ZB semiconductors with SOC taken into account~\cite{Kim-PRB09}. These hybrid functionals are constructed by replacing a quarter of the density functional contribution (in our case the Perdew-Burke-Ernzerhof (PBE) functional~\cite{PBE}) short-range exchange with its Hartree-Fock counterpart, leaving the long-range part unchanged. In the most common HSE06 scheme, the separation into long- and short-range parts is defined by the screening parameter $\mu=0.2$~\AA${}^{-1}$. In this work, however, we use the value $\mu=0.23$~\AA${}^{-1}$, which was reported in Ref.~\onlinecite{Kim-PRB09} to fit the band gap of InSb to its experimental value. That work reported an underestimation by roughly 15$\%$ of the Luttinger parameters obtained from such a calculation compared to experimentally reported values. Here, however, we aim at constructing tight-binding models and the hybrid calculation is an optimal starting point for this purpose.

\begin{table}[h]
\begin{tabular}{|l|c|c|c|c|}
 \hline
 & $E_{\mathrm{g}}$& $\Delta_{\mathrm{SO}}$& $E_{\Gamma_7^\mathrm{c}}-E_{\Gamma_8^\mathrm{v}}$& $E_{\Gamma_8^\mathrm{c}}-E_{\Gamma_8^\mathrm{v}}$\\
 \hline
 $6\times 6\times 6$ & 0.241 & 0.744 & 3.077& 3.488\\
 $8\times 8\times 8$ & 0.236 & 0.746 & 3.070& 3.481\\
 Experiment~\cite{LBInSb} & 0.235& 0.81~\cite{Vurgaftman-JAplP01}& 3.141& 3.533\\
 \hline
 \end{tabular}
\caption{Comparison of our first-principles calculations to experiments. Results for two mesh-densities (6$\times$6$\times$6 and 8$\times$8$\times$8) are given to illustrate convergence. The labels $\mathrm{c}$ and $\mathrm{v}$ refer to conduction and valence bands. All values are given in units of eV.}
 \label{tab:comp}
 \end{table}

First-principles calculations were performed for experimental lattice constants of Ref.~\onlinecite{InSblatt} using projector augmented-wave (PAW) basis sets~\cite{PAW1,PAW2} implemented within the VASP code~\cite{Kresse-PRB96,VASP}. The energy cutoff for the PAW potentials was taken to be $280$~eV. Gaussian smearing of $0.05$~eV width and a $\Gamma$-centered $6\times 6\times 6$ $k$-point mesh were used to perform BZ integrations. A finer $8\times 8\times 8$ mesh was used to check the convergence of our calculations, and we found that it resulted in only a small $\approx 2\%$ (see Tab.~\ref{tab:comp} for actual numbers) decrease of the band gap, leading to an even better match with the experimental value. Overall, we see that the $6\times 6\times 6$ results are sufficiently converged and, given the significant increase in computational cost for finer meshes, we report results for this mesh throughout this paper.

The bonding $p$-states and the antibonding $s$- and $p$-states, forming the topmost valence and lowest conduction bands, are of most interest to us for the present purpose. Ignoring spin-orbit coupling, the $s$-states at the $\Gamma$ point transform according to a one-dimensional representation $\Gamma_1$ of the symmetry group $T_d$, while the $p$-states transform as a three-dimensional representation $\Gamma_{15}$, giving rise to the three-fold degenerate multiplets $\Gamma_{15}^{v}$ and $\Gamma_{15}^{c}$ of valence and conduction states. When SOC is switched on, the bands at $\Gamma$ are classified according to double group representations. The $s$-states now form a $\Gamma_6$ representation that roughly corresponds to the $|S,j=1/2,j_z=\pm1/2\rangle$ states. The $p$-states, which were 6-fold (with spin) degenerate without SOC, now split into a four-dimensional representation $\Gamma_8^{v,c}$, which accommodates the heavy ($|P,j=3/2$, $j_z=\pm 3/2\rangle$) and light ($|P,j=3/2$, $j_z=\pm 1/2\rangle$) holes, and a two-dimensional representation $\Gamma_7^{v,c}$ referred to as the split-off band ($|P,j=1/2,j_z=\pm1/2\rangle$).
\begin{figure}[tb]
\begin{center}
\includegraphics[width=\columnwidth]{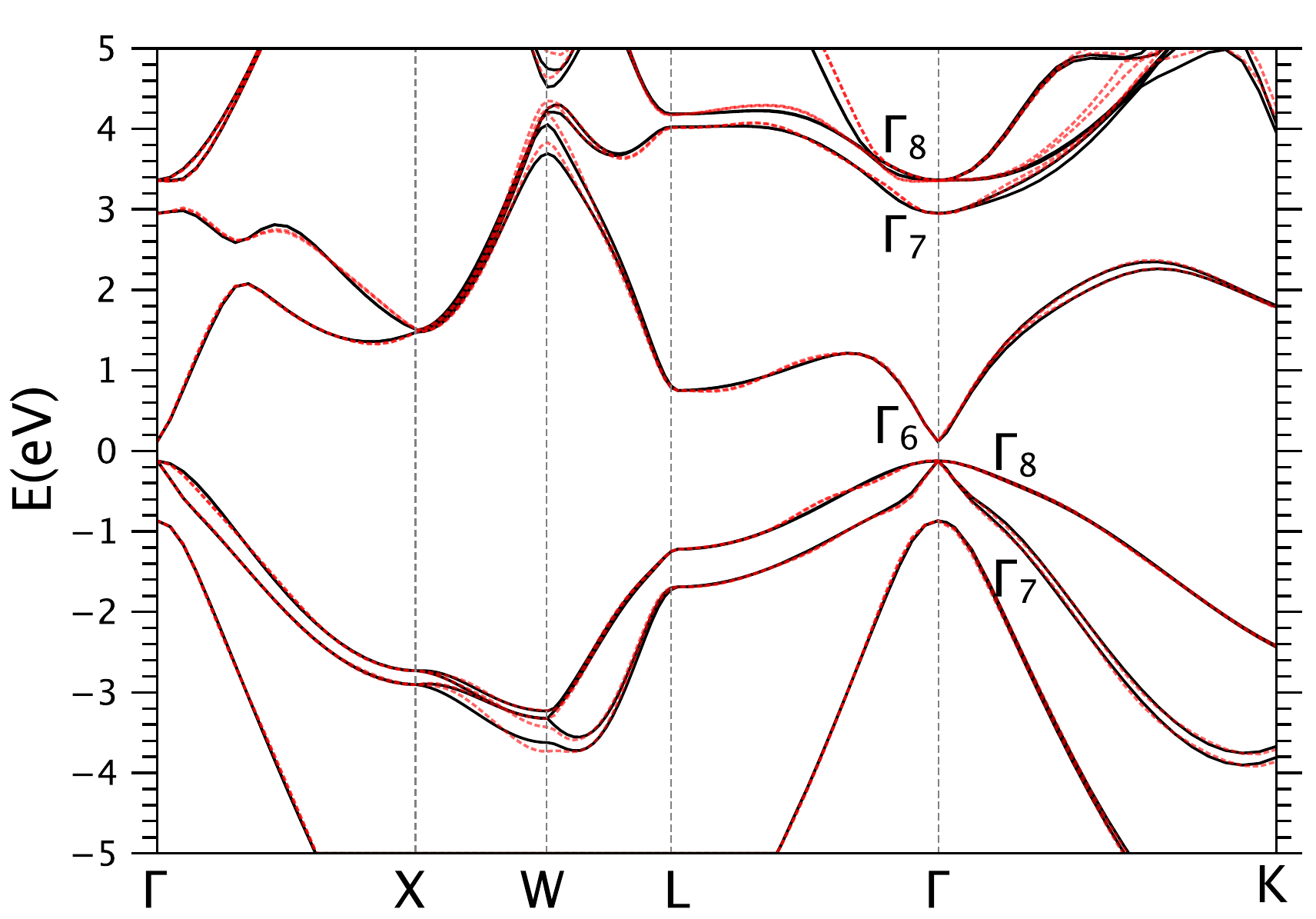}
\end{center}
\caption{Band structure of InSb with SOC. The Fermi level is set at $0$eV and labels of irreducible representations at the $\Gamma$ point are also shown. Black solid lines are the {\it ab initio} band structure. Red dashed lines are the band structure obtained from a tight-binding model of Sec.~\ref{sec:tb}.}
\label{fig:InSbbsso}
\end{figure}
The band structure with SOC taken into account is shown in Fig.~\ref{fig:InSbbsso}, and exhibits the correct band ordering in the BZ center. The results are in excellent agreement with experiments as shown in Table~\ref{tab:comp}.

The bulk band gap $E_{\mathrm{g}}$ is only within $\approx 2.5\%$ of the experimental value~\cite{LBInSb}. Similar agreement is seen for other experimentally known energy differences, with the only exception being the spin-orbit splitting of hole states $\Delta_{\mathrm{SO}}=E_{\Gamma_8^{\mathrm{v}}}-E_{\Gamma_7^{\mathrm{v}}}$, which is only within $\approx 8\%$ of its generally accepted value~\cite{Vurgaftman-JAplP01}. We noticed that by decreasing the mixing parameter $\mu$, we can obtain better agreement for $\Delta_{\mathrm{SO}}$, but at the price of increasing the value of the fundamental gap $E_{\mathrm{g}}$.

The effective mass that we obtain from this calculation is $m^*\approx 0.014 m_e$, where $m_e$ is the electron mass. This value is within the range of reported experimental values~\cite{Vurgaftman-JAplP01} that is from $0.012m_e$ to $0.015m_e$, and agrees perfectly with the generally accepted value of $0.0135 m_e$, proving the reliability of our first-principles calculation. In the following we will use this calculation to construct TB models to study finite-size effects on the spin splittings of the first conduction band $\Gamma_6$.

\section{Tight-binding Models}
\label{sec:tb}

{\it Ab initio} simulations with hybrid functionals are computationally very demanding, which renders the direct {\it ab initio} simulation of wires and large heterostructures unfeasible. To perform such simulations one thus needs to employ tight-binding (TB) models, which we introduce in this section. Being self-consistent, first-principles calculations generally contain much more information than tight-binding models. However, the tight-binding approximation is sufficient for many purposes. Aiming to describe spin splittings, that is, band structure effects, it is reasonable to assume that a good TB model will reproduce them correctly.

We first discuss the existing TB schemes, one based on fitting to experimental data that is widely used in semiconductor physics, and another applied more commonly in materials modeling based on {\it ab initio} calculations. We then introduce a scheme for constructing TB models that is a hybrid of these two methods, where we aim to describe both the spectrum and the wavefunctions of the material reliably. A TB model for InSb obtained with this hybrid scheme is introduced, which will form the basis for obtaining the results of subsequent sections.

\subsection{Empirical tight-binding models}

Tight-binding modeling of semiconductors is usually done using ETB models, where the symmetry-constrained parameters are fitted to experimental data. A variety of parametrization schemes exists~\cite{Slater-PR54, Chadi-PSS75, Vogl-JPCS83, Jancu-PRB98, Klimeck-Superlat00} and it is straightforward to generate new ETB models by, for example, supplementing the experimental data with the results of first-principles calculations. In the absence of experimental data, the fitting is done solely based on first-principles calculations.

For ZB binary materials several standard ETB schemes have been used. The so-called $sp3s^*$ model~\cite{Vogl-JPCS83} contains cation and anion $s$- and $p$-states, plus one additional $s$-state. This model fits the band structure around the $\Gamma$-point reasonably well, giving a good description of the hole bands. The electronic bands tend to have too large effective masses, and away from the $\Gamma$-point the bands tend to flatten out. To avoid these problems, the $sp3d5s^*$ scheme~\cite{Jancu-PRB98} can be used, which includes $d$-states and allows for a better description of both electron and hole bands throughout the BZ.

ETB models are simple since they contain only local and nearest neighbor hoppings. This makes them ideally suited to simulate the effects of disorder~\cite{Shtinkov-PRB03, Hass-Vacuum83, Chadi-PRB77}, and they are the models of choice for many semiconductor simulations. However, fitting the band structure well is not the only requirement for a TB model. In particular, a good TB should reproduce not only the energy spectrum, but also other observables correctly.
Since the calculation of observables is done using the Bloch eigenstates of the TB model, a natural requirement for the TB model is to represent the main features of the wavefunction of electrons in the material correctly. 

Since ETB models are based on fitting the spectrum only, and since there are many different fits that will reasonably reproduce the band structure but not the wave function, ETB models do not necessarily satisfy this requirement. Models that incorporate the correct character of the electronic wave functions can however be derived from {\it ab initio} simulations.

\subsection{Tight-binding models based on Wannier functions}
Here we describe a method used to construct interpolated band structures and to extract tight-binding parameters from {\it ab initio} calculations. This method is based on the works of Refs.~\onlinecite{Marzari-PRB97,Souza-PRB01} and is implemented numerically in the Wannier90~\cite{Wannier90} software package. Rooted in the construction of Wannier functions (WFs) from the Bloch states obtained from first-principles calculations, this method guarantees that the resultant TB model has the correct wavefunction character.

We briefly review the concept of maximally localized WFs~\cite{Marzari-PRB97, Marzari-RMP12}, focusing on the extraction of TB parameters. The problem is stated like this: given a band structure obtained by first-principles calculation, construct a TB model that correctly describes the band structure and wavefunction character for a set of bands that fall within some chosen energy window, referred to as the outer window.

At each ${\bf k}$-point, a certain number $N_{\bf k}\geq N$ of bands, where $N$ is the number of bands of the desired TB model, falls into the outer window. The procedure of Ref.~\onlinecite{Souza-PRB01} is initiated by a guess for the $N$ orbitals (or some localized states) $g_n({\bf r})$ that dominate the character of these bands. These $N$ orbitals are then projected onto the $N_{\bf k}$ bands, forming a set of of non-orthonormal states
\begin{equation}
|\phi_{n{\bf k}}\rangle = \sum_{m=1}^{N_{\bf k}} A_{mn}|\psi_{m{\bf k}}\rangle.
\end{equation}
where $A_{mn}({\bf k})=\langle \psi_{m{\bf k}}|g_n\rangle$ is an $N_{\bf k}\times N$ matrix.

If the set of orbitals was well-chosen, the matrix $S_{mn}=\langle \phi_{m{\bf k}}|\phi_{n{\bf k}}\rangle$ is invertible and a new set of orthonormal Bloch states can be obtained by L\"owdin orthonormalization
\begin{equation}
|\psi^{(0)}_{n{\bf k}}\rangle = \sum_{m=1}^{N} \left( S^{-1/2}\right)_{mn}|\phi_{m{\bf k}}\rangle.
\end{equation}
Using the lattice-periodic parts of these $N$ states $u^{(0)}_{n{\bf k}}=e^{-i{\bf k}\cdot{\bf r}}\psi^{(0)}_{n{\bf k}}$ one further disentangles them by minimizing the spread functional
\begin{equation}
{\cal F}=\frac{1}{N_{kp}}\sum_{{\bf k},{\bf b}}\omega_{\bf b} \sum_{m=1}^N\left(1-\sum_{n=1}^N|\langle u_{m{\bf k}}|u_{n{\bf k}+{\bf b}}\rangle|^2 \right)
\end{equation}
where $N_{kp}$ is the number of ${\bf k}$-points used for the discretization of the BZ. The set of vectors ${\bf b}$ with weights $\omega_{\bf b}$ is used for finite difference discretization of derivatives in ${\bf k}$-space (see Ref.~\onlinecite{Souza-PRB01} for details). Minimization is done self-consistently on the whole ${\bf k}$-mesh with respect to possible choices of $N$ representatives out of $N_{\bf k}$ states~\cite{Souza-PRB01}.

In practice, one often wants to guarantee the presence of some band character at certain $k$-points in the model. For this reason a second (inner) energy window is chosen within the outer window. This inner window contains the features of interest and at each ${\bf k}$-point encloses $M_{\bf k}$ bands. In that case the above disentanglement procedure is done for $N_{\bf k}-M_{\bf k}$ bands only.

The minimization of the spread functional is most intuitive in 1D. The functional is obviously minimized when at each consecutive $k$-point a set of $N$ states is chosen to maximize the overlap with the set of states from the previous $k$ point. Thus, minimizing ${\cal F}$ leads to the smoothest possible choice of $N$ $u_{n{\bf k}}$-functions to represent the states of interest within the chosen energy window.

Once $N$ Bloch states are disentangled from the rest of the spectrum, there remains freedom to rotate these states at each ${\bf k}$-point by a unitary (gauge) transformation $\mathrm{U}_{\bf k}(N)$ to obtain the smoothest possible Bloch states that span the same Hilbert space, as the initial ones. The smoother the Bloch states, the better localized WFs are obtained after Fourier transforming these states. The optimization is done by minimizing the total real space spread of the WFs, producing so-called maximally localized WFs~\cite{Marzari-PRB97}. While for trivial band topologies, these WFs are exponentially localized~\cite{Marzari-PRB97}, a non-trivial band structure can present obstructions to obtaining exponentially localized WFs~\cite{Soluyanov-PRB11-a, Winkler-arx15} and special care must be taken to handle these cases.

After a set of smooth Bloch states, and hence, exponentially localized WFs are obtained, the Hamiltonian matrix is calculated at each ${\bf k}$-point within the chosen subspace of $N$ states~\cite{Souza-PRB01}. The TB Hamiltonian becomes
\begin{equation}
H_{nm}({\bf R}) = \langle w_n({\bf 0})|H|w_m({\bf R})\rangle
\end{equation}
where $w_n({\bf R})$ is the WF obtained as a result of the above procedure, located in the unit cell indexed by a lattice vector ${\bf R}$.

Unlike ETB, thus obtained TB models usually have hoppings to distant neighbors, although their magnitude decreases for well-localized WFs. Such models are used extensively in condensed matter physics~\cite{Marzari-RMP12}. However, they have a problem -- the WFs do not necessarily have the symmetry of the chemical orbitals, or of the basis functions of the corresponding irreducible representations, leading to small symmetry breaking, which results in visible absence of symmetry-protected degeneracies in the TB Hamiltonian.

Being small (typically of order of few tens of meV for a good model), this symmetry breaking can be neglected for many applications. But when dealing with subtle low energy scale effects, like the spin splittings we study in the present paper, the magnitude of the symmetry breaking is often larger than the effects we are after. For this reason, additional symmetrization of the WFs is required.

\subsection{Symmetric tight-binding models}

The symmetry of the WFs in the above procedure already gets broken in the process of disentangling the bands of interest from the rest of the spectrum. As a first indication of this symmetry breaking the charge centers of WFs
\begin{equation}
{\bf r}_n=\langle w_n({\bf 0})|\hat{\bf r}|w_m({\bf 0})\rangle
\end{equation}
usually shift away from the positions of the orbitals that are used for the initial projection. As a result, the Wannier centers form a lattice that is different from the original one, usually leading to breaking of lattice symmetries in the resultant TB model.

Fixing the Wannier centers on the atomic sites improves the symmetry in some cases, as was also mentioned in the independent work of Ref.~\onlinecite{Wang-PRB14}, which suggests to use Lagrange multipliers to enforce fixed Wannier charge center position in the minimization scheme. We notice, however, that the centers shift even before maximal localization, and thus including such Lagrange multipliers can prevent the algorithm from finding optimally localized Wannier orbitals in such cases. Moreover, fixing the symmetric positions of the Wannier centers does not guarantee that the resultant TB model becomes symmetric.

Here we solve this problem by identifying an outer energy window for which the centers do not move as a result of disentanglement (see Appendix~\ref{sec:tbparameters} for details). This choice of the window also works successfully for other binary ZB semiconductors. To obtain better similarity between the WFs and chemical orbitals the symmetry of the hopping parameters should also be taken into account. While Lagrange multipliers can also be used to force hopping terms that violate the symmetry to vanish, we find that using our approach of appropriately tuning the outer energy window remedies this problem as well. In particular, the inter-orbital on-site matrix elements that break the symmetry become small and the hoppings within a unit cell are as expected for the orbitals in the crystal field.

While the TB model used in this paper was obtained by manipulations with energy windows, Lagrange multipliers can in principle be introduced when minimizing the spread of the WFs not only to fix the centers of the WFs, but also to force the hoppings that would vanish by symmetry in the case of chemical orbitals, be zero. Including more and more distant neighbors will eventually result in WFs that represent very good approximations to chemical orbitals. It should be stressed that in order to obtain better localization of the symmetric WFs it is still necessary to find an energy window, in which the symmetry conditions are not broken too strongly. Besides, this approach can only work, provided the initial energy window and projection capture the necessary orbital character throughout the BZ.

From this procedure we obtain TB models that have two desired properties: they correctly reproduce the wave functions and are written in the basis that is close to that of atomic orbitals. If such TB models are created based on {\it ab initio} simulations separately for the bulk materials, and for interfaces/surfaces using superlattices, then by gluing them together one can model realistic devices and heterostructures.

Another benefit of this type of models is that they can be constructed without SOC taken into account, which in many cases saves an enormous amount of computational effort. If the resultant WFs are orbital-like, being well localized, it is reasonable to approximate the effect of SOC as an on-site term, constrained by symmetry. The form of the local SOC for $p$-, $d$- and $f$-orbitals is known~\cite{Friedel-JPCS64, Chadi-PRB77, Jones-PRB09}, and each of them has only one parameter that can be fitted to either experimental or first-principles data. This fitting is in the flavor of ETB but has much less free parameters and starts with the correct wave function behavior. Finally, having the correct wave function behavior, these models are ideally suited for calculations of $g$-factors and finite-size induced spin splittings, being potentially very useful for a wide range of applications.

The TB model for InSb obtained according to the above procedure is detailed in Appendix~\ref{sec:tbparameters}. The effective mass obtained with this model is $m^*\approx 0.015 m_e$, changing slightly from the its first principles value. For the purposes of the present paper we used only $s$- and $p$-states to create the TB model, and implemented local approximation (see Appendix~\ref{sec:tbparameters}) to the SOC. In ETB models this approximation in $sp3s^*$ models is known to miss small linear in $k$ splittings in the hole bands~\cite{Boykin-PRB98} due to omission of the $d$-states, but the description of the electron bands that are of interest in this work is still reliable.

\section{Spin-orbit splittings}
\label{sec:so}

In this section we discuss induced spin splittings in InSb thin films. For this purpose slabs of various thicknesses and orientations are considered. This allows for the evaluation of the finite size effects on the spin splitting (Sec.~\ref{sec:finite}). More importantly, the discussion of the splittings induced by applying an external electric field in the direction, orthogonal to the slab is presented (Sec.~\ref{sec:fieldso}). This allows to identify both BIA and SIA contributions to the spin splittings in the slab geometry and fit them to the analytic symmetry-based models. In a wire both these contributions combine to give the effective splitting
\begin{equation}
\Delta E= \frac{m^*{\alpha^*}^2}{2\hbar^2}
\label{eq:de}
\end{equation}
where $\alpha^*$ is an effective 1D Rashba parameter. Wire directions, for which this splitting and its susceptibility to the external electric field  is maximized are also identified.

\begin{figure}
\begin{center}
\includegraphics[width=\columnwidth]{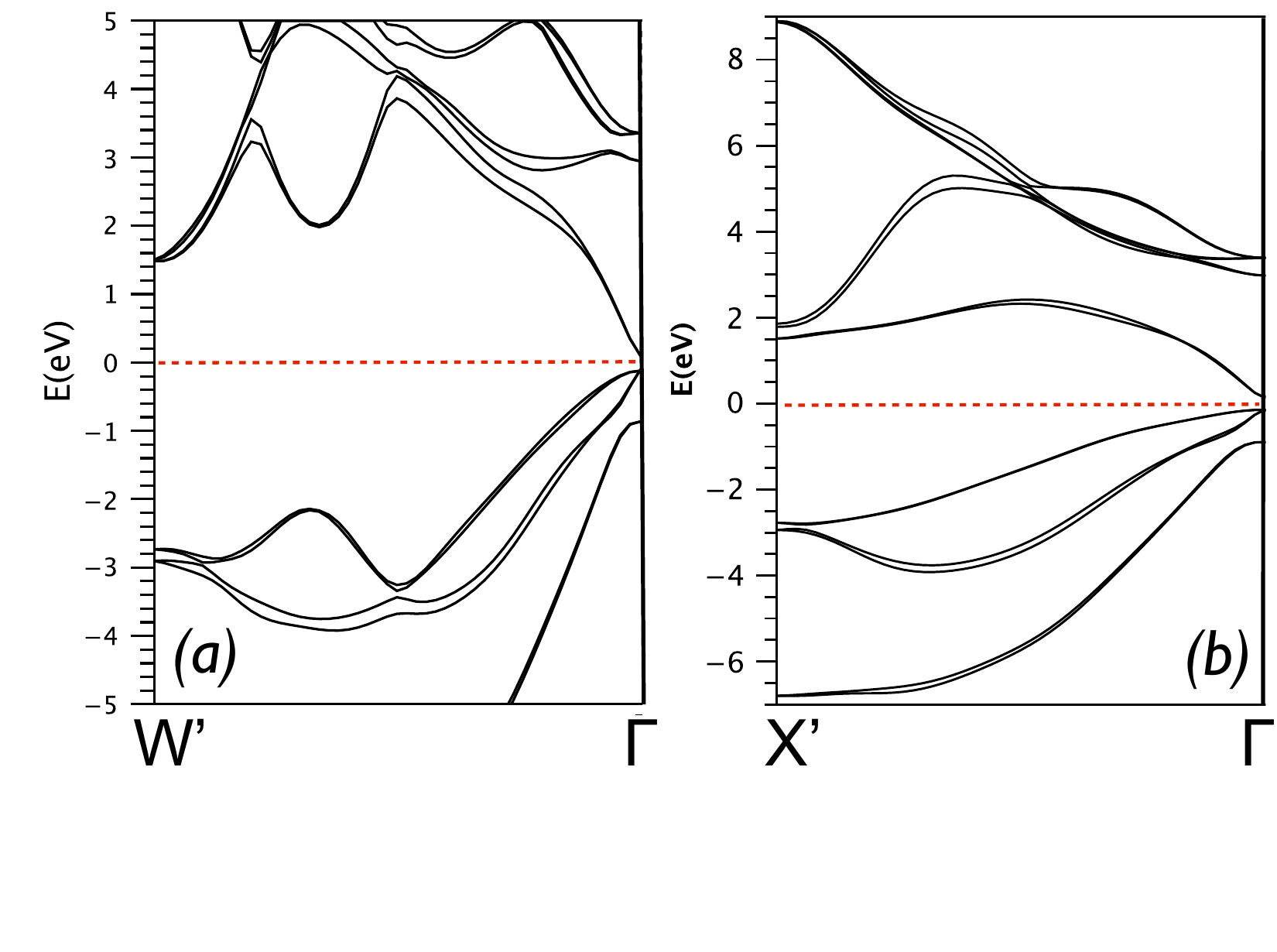}
\end{center}
\caption{Bulk band structure of InSb along the (b) $[210]$-direction and (b) $[110]$-direction. The largest spin splitting of the conduction band occurs along the $[210]$-direction.}
\label{fig:dress}
\end{figure}

In the seminal work of Dresselhaus~\cite{Dresselhaus-PhysRev55} it was shown that no linear in $k$ spin splitting terms appear for the electronic band in the vicinity of the BZ center. The splitting is cubic in $k$ and the effective Hamiltonian for the electronic band can be written as~\cite{Ganichev-PSS14}
\begin{equation}
H_{\mathrm{ZB}}({\bf k})=\frac{\hbar^2 k^2}{2 m^*} +H_{\mathrm{S}}
\end{equation}
where
\begin{equation}
H_{\mathrm{S}}=\gamma \left[ k_x(k_y^2-k_z^2)\sigma_x+k_y(k_z^2-k_y^2)\sigma_y+k_z(k_x^2-k_y^2)\sigma_z \right]
\label{eq:dress}
\end{equation}
and $\gamma=760$~eV/\AA$^{-3}$ is the generally accepted value~\cite{Winkler-book}. This expression predicts the largest splitting for the bulk conduction band to appear in the $[110]$ direction. However, this conclusion holds only for small momenta. For larger $k$ the largest splitting is found to occur in the $[210]$ direction, as illustrated in Fig.~\ref{fig:dress}. Similar findings were previously reported in Ref.~\onlinecite{Luo-PRL09} for GaAs and GaSb.

\subsection{Method}

Inclusion of an electric field into the bulk calculation is a tedious task~\cite{Souza-PRL02}, since the electrostatic potential does not have the lattice periodicity and the translation invariance is broken along the field direction. The consideration is significantly simplified in the case of a slab calculation, when the electric field is applied orthogonal to the slab, as illustrated in Fig.~\ref{fig:slab}(b). The two in-plane momenta are still good quantum numbers and the field-induced spin splittings are thus most easily accessible in a slab calculation.

Since {\em ab initio} simulations using hybrid functionals are limited to very small systems, TB models of Section~\ref{sec:tb} need to be used to determine the spin splittings in the presence of an electric field. We consider two different slabs: $(001)$ and $(110)$ ones, which are perpendicular to the $[001]$ and $[110]$ directions correspondingly. To double check our conclusions for the $(110)$-slab, we also considered a symmetrically equivalent slab $(1\bar{1}0)$, orthogonal to the $[1\bar{1}0]$-direction.

The effect of electrostatic potential is approximated by a contribution to the on-site potential. This approximation, generally accepted in the TB modeling~\cite{Graf-PRB95}, requires some justification here. Unlike the usual TB, where the orbitals are assumed to be very strongly localized, our TB model is based on the WFs that have a finite spread. However, as clarified in the Appendix~\ref{sec:tbparameters}, these WFs are also well localized and their spread is smaller than interatomic distance, so that the local approximation for the electrostatic potential still holds. 

\subsection{Finite size splittings}
\label{sec:finite}

The surfaces of the slabs cause complications that need to be properly dealt with. Real material surfaces and interfaces can be very complex due to lattice reconstruction, local strains and disorder. These effects are not captured by the present TB model. While some attempts to model such effects from-first-principles can be done for GaAs and AlAs, where GGA still produces the correct band ordering, the use of hybrid functionals needed for such a calculation for InSb is extremely demanding computationally. For the purpose of this paper we are, however, not interested in a microscopic description of surfaces and interfaces but rather in bulk effects at a safe distance from any surface or interface. Changes of the crystal structure at the surface/interface create an effective intrinsic (electric) field within the bulk of the slab, and the magnitude of this field depends on the design of the heterostructure. Also, in the case of epitaxially grown devices~\cite{Krogstrup-NatMat15} the effects of interface lattice reconstruction and interface disorder are minimized, so that the intrinsic field strength is solely determined by the SIA of the heterostructure and its material composition. In what follows we analyze structures that are much thicker than the region potentially influenced by the changes in the lattice structure at the surface. This is verified by checking that the wavefunction of the lowest conduction band is localized in the bulk of the material of interest, quickly decaying towards the surface. For this reason in our study it is sufficient to only study the effects of intrinsic electric fields on the spin-splittings of the first conduction band.

We thus approximate the surface by truncating all hoppings into the vacuum region. The truncation, however, generates in-gap surface states due to unsaturated (dangling) covalent bonds for certain surface orientations. Since InSb is not a topological insulator, all in-gap surface states can be avoided by local changes to the Hamiltonian at the surface without much influence on the bulk wavefunction that is of primary interest. In a real material the dangling bond states are usually eliminated by either surface reconstruction or by saturation with adatoms. The effect of atoms saturating the dangling bond can be introduced in a TB model by passivation. This passivation of the dangling bonds can be done in several ways~\cite{Lee-PRB04}.

\begin{figure}[t]
\begin{center}
\includegraphics[width=\columnwidth]{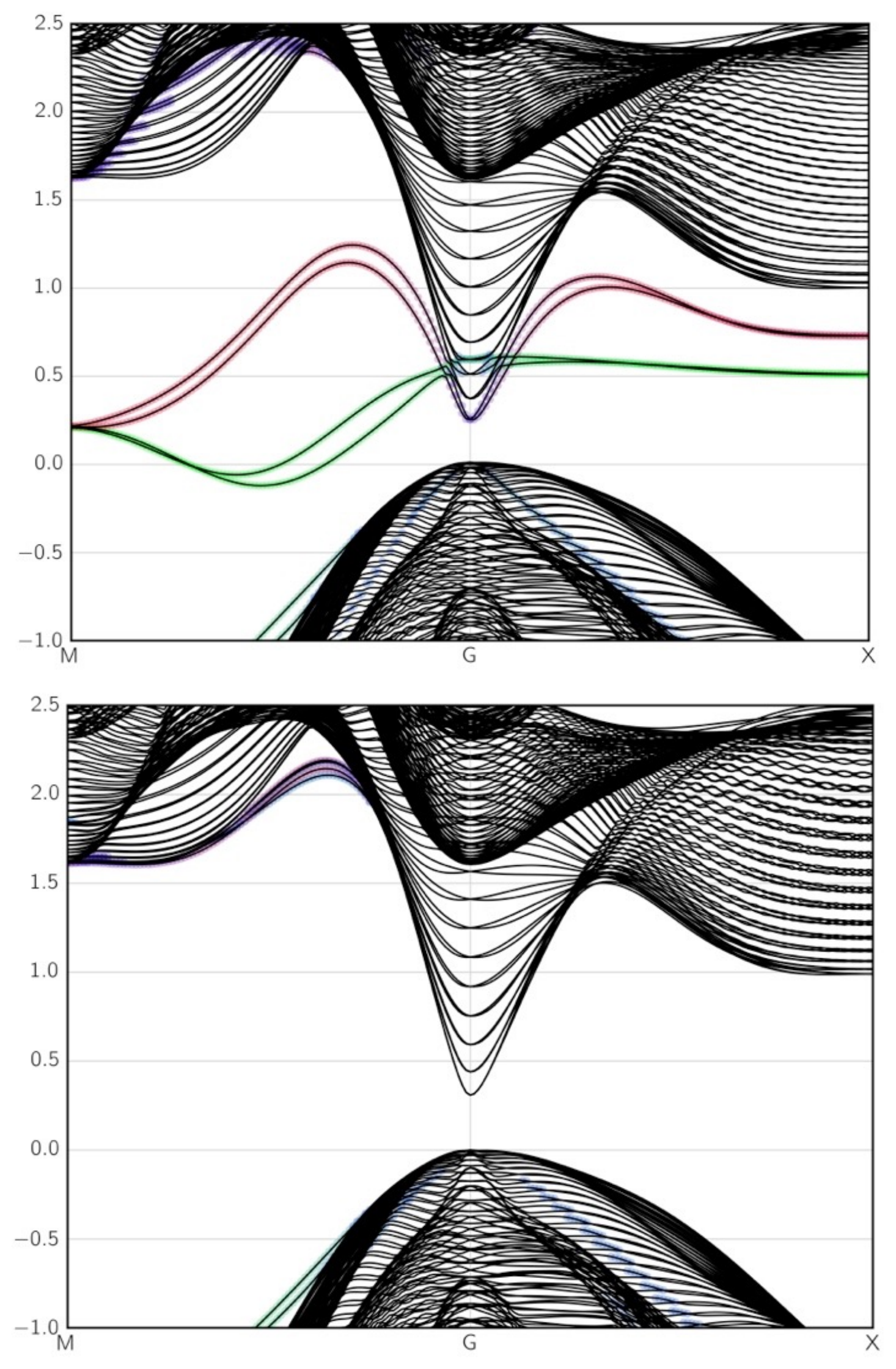}
\end{center}
\caption{Band structure of a $\approx$162\AA-thick $[001]$-slab of InSb. Color scheme describes the contribution of the atoms from the top (red) and bottom (green) surfaces to the weight of the subband wavefunction. Top panel: without passivation. Bottom panel: with passivation. $X$ refers to the point $(\pi,0)$ and $M$ to $(\pi,\pi)$ of the 2D BZ of the slab.}
\label{fig:pass1}
\end{figure}

For the slab orientations considered in this paper, dangling bonds are most visible at the $(001)$-surface, as illustrated in Fig.~\ref{fig:pass1} for a 50 unit cells thick slab. Changing the on-site energy of $s$ and $p_x$ orbitals of In atoms on the bottom surface by $\epsilon_{s}^{(In)}=\epsilon_{p_x}^{(In)}=5$~eV, and the energy of $p_y$ and $p_z$ orbitals of Sb atoms on the top surface by $\epsilon_{p_y}^{(Sb)}=\epsilon_{p_z}^{(Sb)}=-5$~eV removes the dangling bond states from the gap region with negligible influence on the rest of the spectrum. The choice of these orbitals for passivation is dictated by their dominating spectral weight in the dangling bond surface states. For the $(110)$ and $(1\bar{1}0)$ surfaces, the effect of the dangling bonds is not as drastic. Passivation was done for the $(110)$ surface by changing the on-site energy by $\epsilon_s^{(In)}=\epsilon_{p_y}^{(In)}=\epsilon_{p_z}^{(In)}=1.67$~eV and $\epsilon_{p_y}^{(Sb)}=\epsilon_{p_z}^{(Sb)}=-2.5$~eV at both surfaces. Similarly, the on-site energy on both surfaces was changed by $\epsilon_{s}^{(In)}=5$~eV and $\epsilon_{p_x}^{(Sb)}=\epsilon_{p_y}^{(Sb)}=\epsilon_{p_z}^{(Sb)}=-1.67$~eV for the $(1\bar{1}0)$ slab.

\begin{figure}
\begin{center}
\includegraphics[width=0.9\columnwidth]{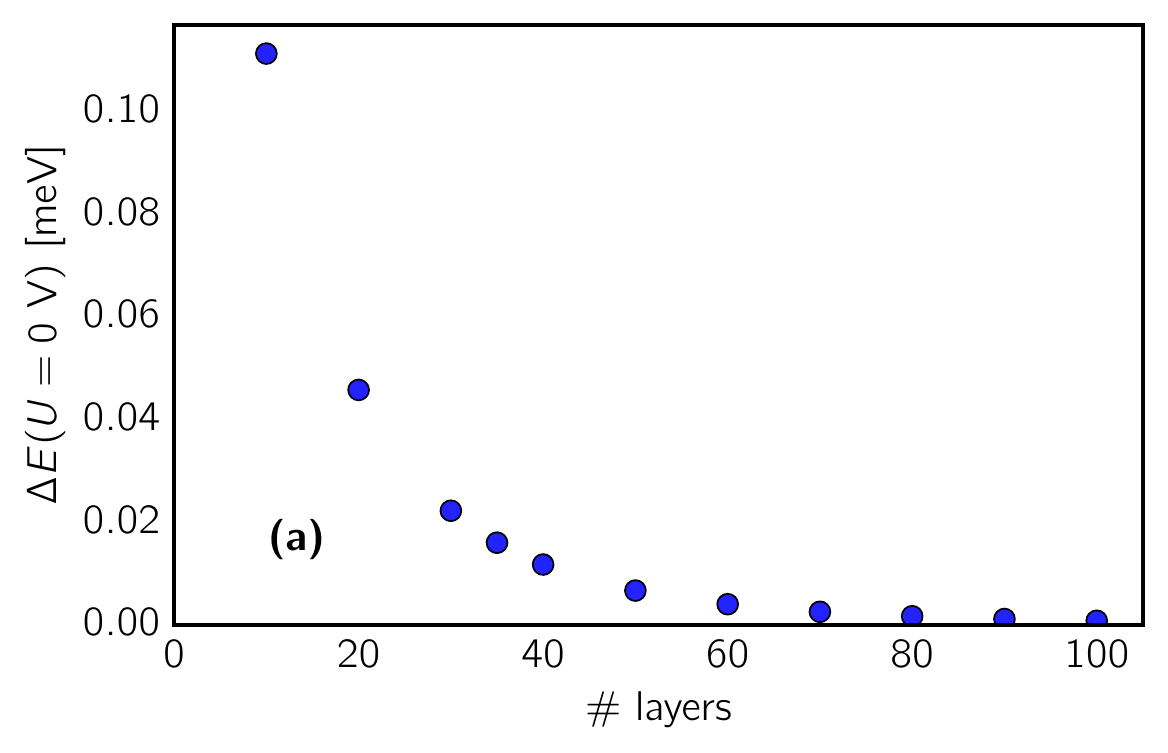}
\includegraphics[width=0.9\columnwidth]{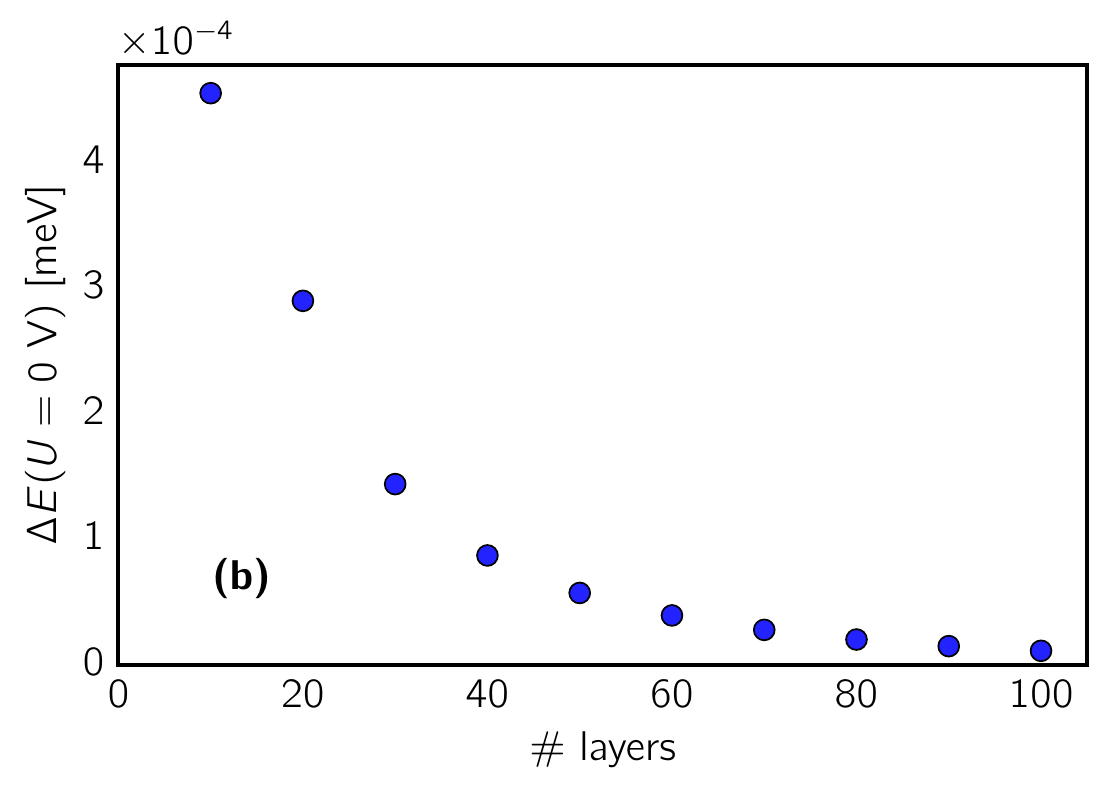}
\includegraphics[width=0.9\columnwidth]{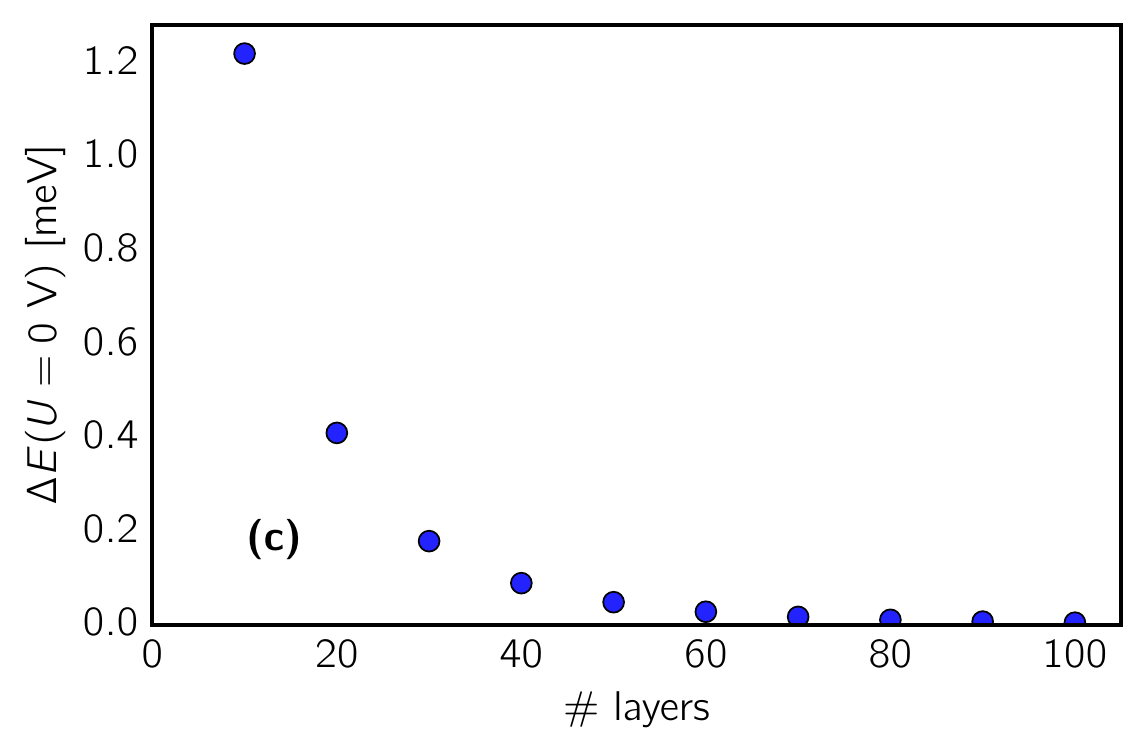}
\end{center}
\caption{Scaling of surface-induced (no external electric field) part of the spin splittings $\Delta E_0$ along the $k_x||[100]$ direction with slab thickness. (a) $[001]$-slab. (b) $[110]$-slab. (c) $[1\bar{1}0]$-slab. In each case, the zero-field contribution decays with increasing number of layers.}
\label{fig:e0_nscaling}
\end{figure}

The illustration of Fig.~\ref{fig:e0_nscaling} shows the finite-size-induced spin splitting with the above described passivations. The surface-induced effect clearly disappears with growing slab widths. Since the wires used in Majorana experiments are typically thick (of order $50-100$nm~\cite{Mourik2012}), we neglect this splitting in the following.

\subsection{Field-induced spin-orbit splitting}
\label{sec:fieldso}

Applying the electric field to TB-modeled slab allows one to scan the band structure in momentum directions perpendicular to the field. The calculations were done for the slabs of $35$ unit cell thickness ( $\approx 115$~\AA) for the $(001)$ slab, and of $50$ unit cells ($\approx113$\AA) for the $(110)$ and $(1\bar{1}0)$ slabs. As discussed above, the finite size induced SO splitting becomes negligibly small at these slab thicknesses and, therefore, we concentrate on the electric field induced contribution, subtracting the finite-size induced contribution. That is, $\Delta E= \Delta E_U-\Delta E_0$, where $\Delta E_U$ is the full splitting seen in the slab subject to the potential difference $U$, and $\Delta E_0$ is the finite-size (zero external field) contribution. We note that for each slab direction the application of an external potential leads to a decrease of the band gap, and a critical value of $U$ exists, for which the band gap closes.

We induce a electric field in the nanostructure by applying a potential difference of $0.2$ or $0.4$~V between the upper and lower surface of the slab. The effects of screening are not taken into account here, and the electric field inside the slab is simply given by the voltage difference divided by the thickness $d$. The corresponding electric field strengths $|\cal {E}|$ are $1.75$~mV/{\AA} and $3.5$~mV/{\AA} for the three directions. We now proceed to the detailed analysis of the numerical results.

\subsubsection{$(001)$-slab}

The crystal structure of this slab has the point group symmetry $D_{2d}$. When the electric field is added orthogonal to the slab the symmetry reduces to $C_{2v}$, which consists of two mirror symmetries and a $C_2$-rotation, which is the product of the mirrors. Taking the $x$ ($y$) axis in the slab to be along the $[100]$ ($[010]$) direction, as illustrated in Fig.~\ref{fig:slab}(a), the mirror symmetries are the ones taking $M_1:\, (x,y)\rightarrow (y,x)$ and $M_2:\,(x,y)\rightarrow (-y,-x)$. In addition to point group symmetries, time-reversal should be taken into account. The spin components $s_i=(\hbar/2) \sigma_i$ transform according to $C_2:\, (s_x,s_y,s_z)\rightarrow(-s_x,-s_y,s_z)$, $M_1:\, (s_x,s_y,s_z)\rightarrow (-s_y,-s_x,s_z)$ and $M_2:\, (s_x,s_y,s_z)\rightarrow(s_y,s_x,s_z)$. Taking into account that spin flips under time-reversal, and that momentum ${\bf k}$ transforms as a vector, the linear in $k$ spin spitting part of the Hamiltonian can be uniquely determined. The resultant Hamiltonian for the conduction band in the vicinity of the $\Gamma$ point to the second order in ${\bf k}$ is~\cite{Ganichev-PSS14}
\begin{equation}
H_{(001)}=\frac{\hbar^2 (k_x^2+k_y^2)}{2m^*}+\alpha(k_x\sigma_y-k_y\sigma_x)+\beta(k_y\sigma_y-k_x\sigma_x)
\label{h001}
\end{equation}
where $k_x$ and $k_y$ correspond to the $\langle10\rangle$ and $\langle01\rangle$ directions in the reciprocal space of the slab. The first spin splitting term, called the Rashba term~\cite{Bychkov-JETPL84}, describes SIA. The second spin splitting term is the Dresselhaus term describing BIA~\cite{Winkler-book, Ganichev-PSS14}. The Rashba coefficient $\alpha$ can be manipulated by electric fields~\cite{Ganichev-PRL04}, while the coefficient $\beta$, according to the Eq.~\ref{eq:dress} is (assuming only linear in ${\bf k}$ terms in the spin splitting) $\beta=\gamma \langle k_z^2\rangle$, where the average is taken with respect to the full 3-dimensional lowest electronic subband wavefunction.

The dispersion is given by
\begin{equation}
\varepsilon_\pm=\frac{\hbar^2 (k_x^2+k_y^2)}{2m^*}\pm\sqrt{(\alpha k_y+\beta k_x)^2+(\alpha k_x+\beta k_y)^2}
\end{equation}
Setting $k_x=k\cos{\phi}$ and $k_y=k\sin{\phi}$ the analytic expression for the spin splitting can be obtained
\begin{equation}
\Delta E = \frac{m^*}{2\hbar^2}(\alpha^2+\beta^2+2\alpha \beta \sin{2\theta})
\label{eq:de1}
\end{equation}
This expression can be fitted to the results of the TB calculation illustrated in Fig.~\ref{fig:001} to give numerical estimates for $\alpha$ and $\beta$. For this estimation we subtract the surface contribution $\Delta E_0$ to $\Delta E$, since passivation used to saturate the surface dangling bonds is not a good approximation for the realistic surface effects. The results are summarized in Tab.~\ref{tab:alpha_beta}. It can be seen that not only $\alpha$, but $\beta$ as well is influenced by the electric field. This can be expected since the full 3D-wavefunction is changed by the presence of the electric field. Moreover, the dependence of $\Delta E$ on the direction within the slab is apparent in Fig.~\ref{fig:001}: the splitting for the $[110]$ direction is twice larger than that in the $[-110]$ direction. This result is especially important for gate-defined nanowires~\cite{Reuther-PRX13,Shabani-arx14} in two-dimensional electron systems. The optimal direction for such wires created in $(001)$ thin films is $[110]$.

\begin{table}[h]
\begin{tabular}{|c|c|c|c|c|}
\hline
$\cal{E}$~[meV /\AA] & & 0 & 1.75 & 3.5\\\hline\hline
& [meV  \AA] & &&\\\cline{2-5}
$(001)$& $|\alpha|$ & $108$ & $385$ & $730$\\\cline{2-5}
&$|\beta|$ & $20.4$ & $120$ & $139$\\\hline\hline

& [meV$^2$  \AA$^2$] &&&\\\cline{2-5}
$(110)$& $\alpha_1^2$ & $58.4$ & $1.81 \times 10^5$ & $6.41 \times 10^5$\\\cline{2-5}
&$\alpha_2^2 + \beta^2 - \alpha_1^2$ & $3.54 \times 10^4$ & $1.88 \times 10^4$ & $1.75 \times 10^4$\\\hline\hline

& [meV$^2$  \AA$^2$] &&&\\\cline{2-5}
$(1\bar{1}0)$& $\alpha_1^2$ & $25.2$ & $2.05 \times 10^5$ & $7.51 \times 10^5$\\\cline{2-5}
&$\alpha_2^2 + \beta^2 - \alpha_1^2$ & $7.17 \times 10^4$ & $8.41 \times 10^4$ & $1.15 \times 10^5$ \\\hline
\end{tabular}
\caption{The values of spin splitting parameters obtained from fitting $\Delta E$ of Eqs.~\ref{eq:de1} and~\ref{eq:de2}. The slab thickness is $\approx 115$\AA\,  in all cases.}
\label{tab:alpha_beta}
\end{table}

From the Fig.~\ref{fig:001} it can be seen that the largest spin splitting occurs along the $[110]$ direction, where Dresselhaus and Rashba terms combine to give an effective 1D Hamiltonian
\begin{equation}
H_{[110]}=\frac{\hbar^2k^2}{2m^*}\pm \frac{\alpha^*}{\sqrt{2}} k(\sigma_y-\sigma_x)
\label{h1d001}
\end{equation}
where $\alpha^*=|\alpha |+|\beta |$. This Hamiltonian can be used to describe $[110]$-wires of ZB InSb subject to the $[001]$ electric field. The variation of the crystal potential at the wire interface creates an effective electric field, whose direction can be manipulated by choosing varying the growth direction.

It should be noted, that in the Hamiltonian of Eq.~\ref{h1d001}, obtained from the 2D dispersion, an additional term that appears due to confinement of the wire in 1D is neglected. The motivation for this is the following. As described above, in the absence of an external electric field the splitting decays quickly with increasing wire thickness, and the corresponding Dresselhaus term tends to zero. Application of the electric field significantly modifies the shape of the subband wavefunction in the direction of the field, but in the transverse direction one can expect it to be the same. Thus, one can argue that if the confinement effect induces negligible spin splitting in wide slabs, the confinement of the thick wire can also generate only very small spin splitting. This argument is also supported by the calculations of the Ref.~\cite{Luo-PRB11}, where no linear in momentum spin splitting of the first conduction band was found in $[100]$ and $[111]$ ZB GaAs wires, and the splitting in the $[110]$ direction was found to be small and decreasing with the radius of the wire.

\subsubsection{$(110)$-slab}
\begin{figure}[t]
\begin{center}
\includegraphics[width=\columnwidth]{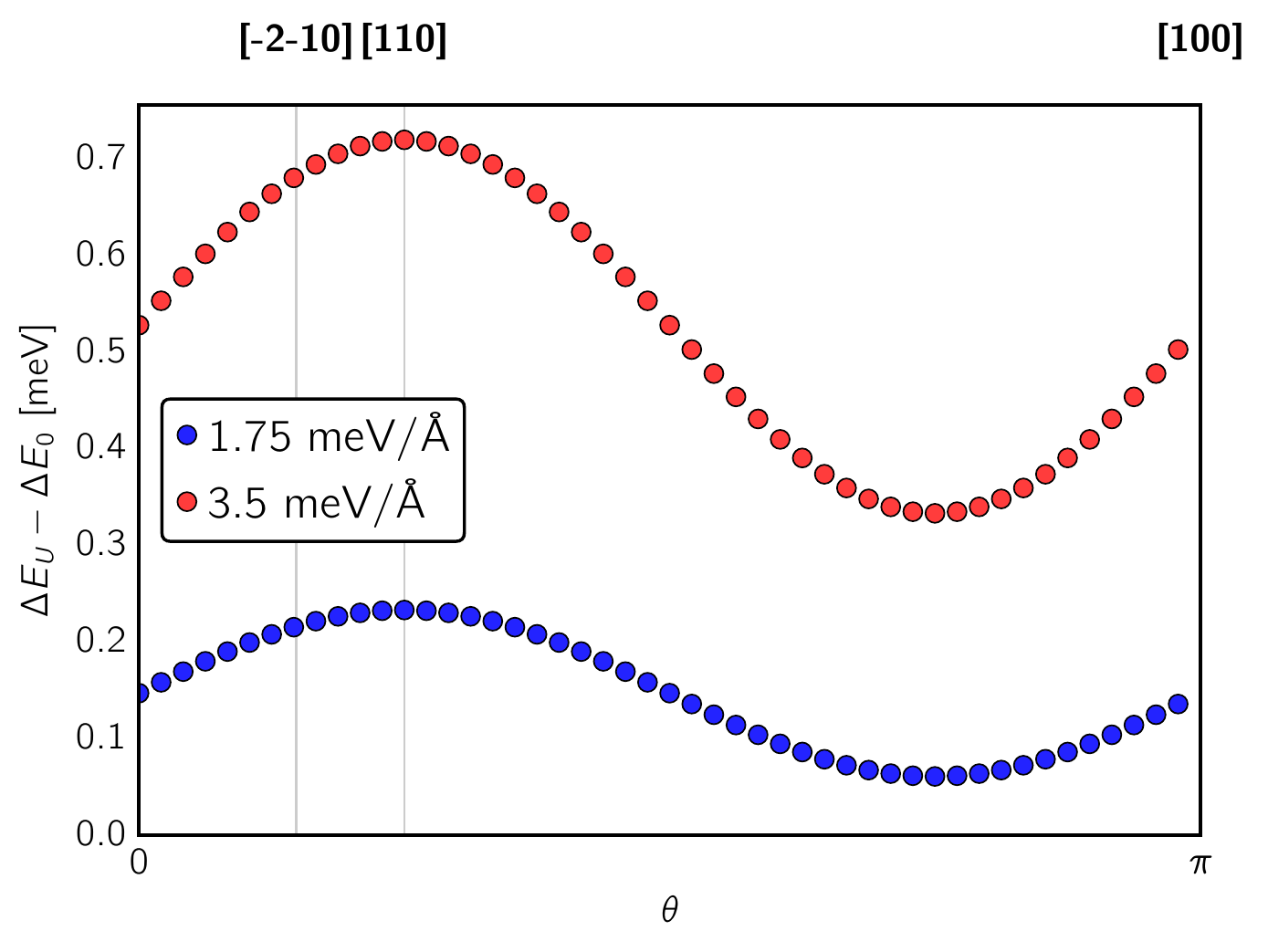}
\end{center}
\caption{Spin splitting $\Delta E - \Delta E_0$ induced by an applied electric field for the $(001)$-slab ($35$ layers, $115$ \AA). The field (blue: $1.75$~mV/\AA; red: $3.5$~mV/\AA) is oriented along $[001]$. $\theta$ of Eq.~\ref{eq:de1} is the angle from the $k_x$-axis in the slab BZ. High-symmetry directions within the slab are marked.}
\label{fig:001}
\end{figure}

\begin{figure}[t]
\begin{center}
\includegraphics[width=\columnwidth]{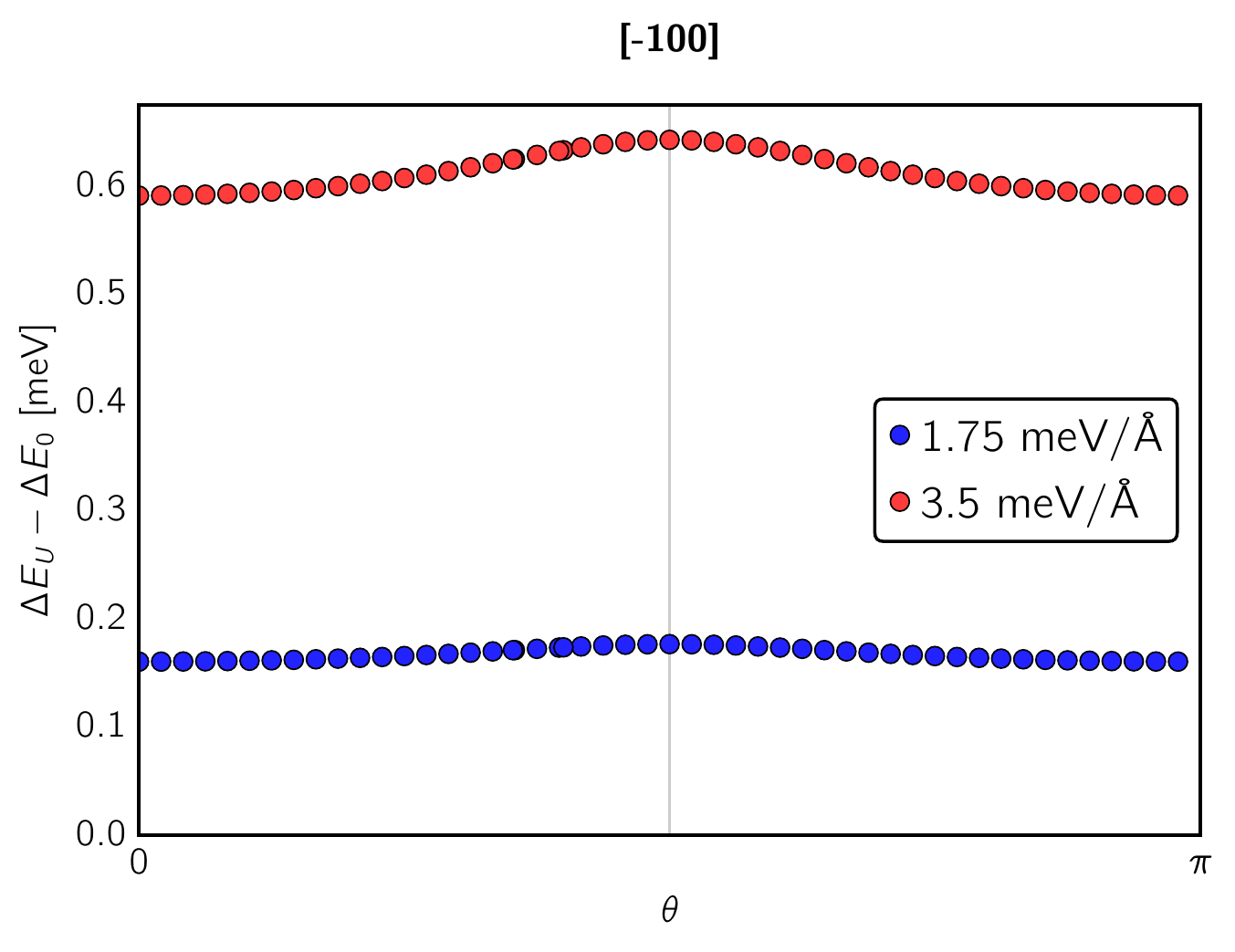}
\end{center}
\caption{Spin splitting $\Delta E - \Delta E_0$ induced by an applied electric field for the $(110)$-slab ($50$ layers, $113$ \AA). The field (blue: $1.75$~mV/\AA; red: $3.5$~mV/\AA) is oriented along $[110]$. $\theta$ of Eq.~\ref{eq:de2} is the angle from $k_x$-axis in the slab BZ. High-symmetry directions within the slab are marked.}
\label{fig:110}
\end{figure}

The point group for this slab~\cite{Nestoklon-PRB12, Ganichev-PSS14} in the presence of external field is $C_s$, with only one mirror plane $(x, y)\rightarrow (-x, y)$, where $x$ is along the $[1\bar{1}0]$-direction of the conventional unit cell and $y$ is along $[001]$. This mirror symmetry takes $(s_x,s_y,s_z)\rightarrow(s_x,-s_y,-s_z)$. The linear in $k$ terms consistent with this symmetry and time-reversal are $k_x\sigma_z$, $k_x\sigma_y$ and $k_y\sigma_x$. The corresponding Hamiltonian is
\begin{equation}
H_{(110)}=\frac{\hbar^2(k_x^2+k_y^2)}{2m^*}+\alpha_1 k_y\sigma_x +\alpha_2 k_x\sigma_y+\beta k_x \sigma_z
\end{equation}
where $\alpha_{1,2}$ now includes also contributions due to the bulk inversion asymmetry, since averaging the bulk splitting of Eq.~\ref{eq:dress} with respect to the 3D-wavefunction gives $\alpha_2=\gamma\langle k_1\rangle/(2\sqrt{2})$, where $k_1$ is along the $[110]$-direction of the conventional unit cell. The corresponding spin splitting is given by
\begin{equation}
\Delta E = \frac{m^*}{2\hbar^2}\left( \alpha_1^2+(\alpha_1^2-\alpha_2^2+\beta^2)\cos^2{\theta}\right)
\label{eq:de2}
\end{equation}
where $\theta$ is the angle in the $(110)$-slab counted from the $[1\bar{1}0]$-direction. The values of $\Delta E$ corresponding to this slab obtained from a TB simulation are illustrated in Fig.~\ref{fig:110}. The fit for $\alpha_{1,2}$ and $\beta$ is given in Tab.~\ref{tab:alpha_beta}. While the angle dependence of $\Delta E$ is less apparent in this case than for the $(001)$ slab, a comparison with the results of Fig.~\ref{fig:001} for the $[-100]$ direction suggests that $\Delta E$ depends on the field direction for the structures in question: for approximately the same slab thicknesses the corresponding splitting in the $[110]$ field is slightly larger than that in the $[001]$ field for the illustrated field strengths.  

\begin{figure}[b]
\begin{center}
\includegraphics[width=\columnwidth]{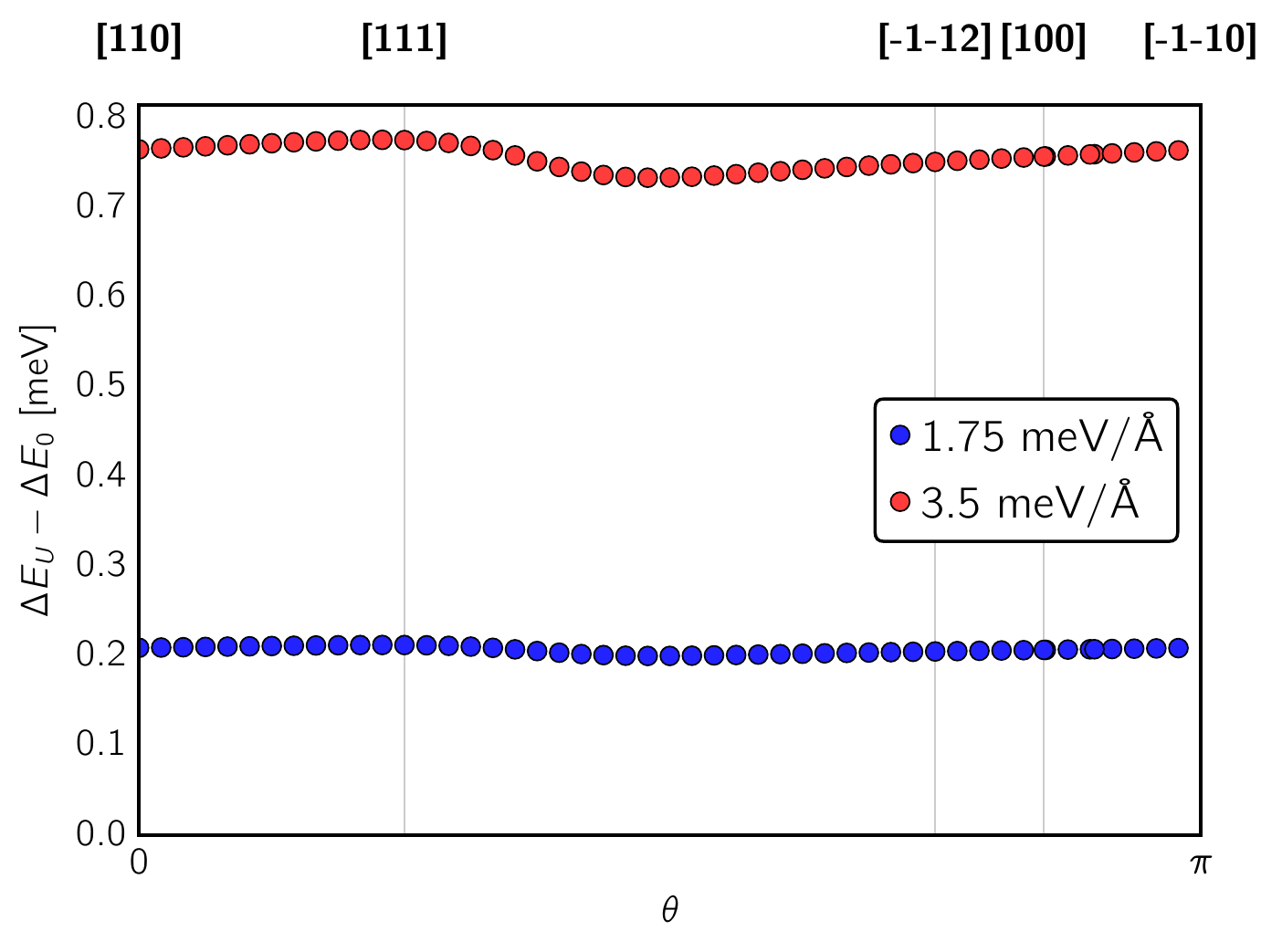}
\end{center}
\caption{Spin splitting $\Delta E - \Delta E_0$ induced by an applied electric field for the $(1\bar{1}0)$-slab ($50$ layers, $113$ \AA). The field (blue: $1.75$~mV/\AA; red: $3.5$~mV/\AA) is oriented along $[1\bar{1}0]$. $\theta$ is the angle from $k_x$-axis in the slab BZ (see Eq.~\ref{eq:de2}). High-symmetry directions within the slab are marked.}
\label{fig:1m10}
\end{figure}

There appears to be little dependence of $\Delta E$ on the angle for this slab direction. To verify this we simulated a symmetry-equivalent $(1\bar{1}0)$ slab. The illustration of the corresponding spin splittings is given in Fig.~\ref{fig:1m10}. Apart from the finite size zero-field effects, mediated by the differences in passivation, this simulation also predicts little variation of $\Delta E$ with $\theta$.

\begin{figure}[t]
\begin{center}
\includegraphics[width=\columnwidth]{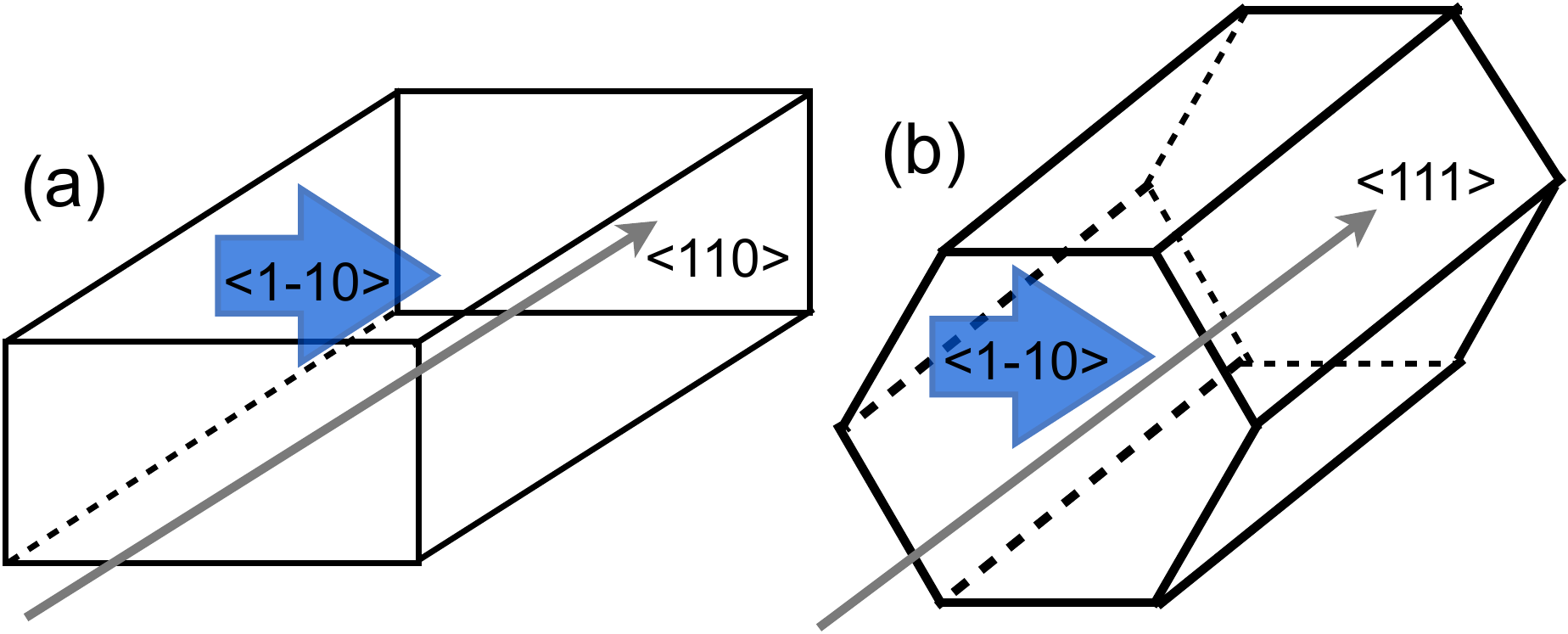}
\end{center}
\caption{Some of the optimal wire/field directions. (a) A $[110]$-wire in in the $[1\bar{1}0]$ field. (b) A $[111]$-wire in in the $[1\bar{1}0]$ field.}
\label{fig:wires}
\end{figure}

The $(1\bar{1}0)$-slab also contains the $[111]$-wire that was used in the original MZM experiment~\cite{Mourik2012}. The data presented in Figs.~\ref{fig:001}-\ref{fig:1m10} indicates that this wire direction should have almost optimal spin splitting (see Fig.~\ref{fig:wires} for the illustration of several optimal wire directions). Magnetoconductance measurement were performed recently~\cite{Weperen-PRB15} to estimate the size of the spin splitting in the $[111]$-wires grown in a setup relevant for MZM. The measured range for the spin splitting is $\Delta E \approx 0.25-1$meV. These values can be compared to the data obtained in the present TB simulation to give an estimate for the strength of electric field inside the wire.

Figure~\ref{fig:Uscaling_2} in Appendix~\ref{sec:scaling} shows that $\Delta E \propto |{\cal E}|^2$ for $|{\cal E}| \rightarrow 0$. The dependence is almost the same for all directions within the $(110)$ slab. Using the effective mass of $m^*=0.015 m_e$ obtained in the TB model and the experimentally reported range of $\Delta E$, one can back out the electric field needed to generate such a SOC: $e|{\cal E}|\approx 2-4$meV/\AA. Thus, the method presented in this work can be used to estimate the size of electric field inside a material, which is important for the experiments trying to control SOC couplings by gating the nanowire.


\section{Conclusions and Outlook}
\label{sec:end}

The problem of optimizing spin splittings in the semiconductor is one of the main ingredients for the possible realization of MZMs in proximity coupled semiconductor nanowires. In this work we addressed this problem using a TB model of InSb derived from the modified version of the highly accurate HSE06 hybrid functional. This allowed for the detailed study of finite-size and field-induced spin splitting in slabs of InSb. These results are used to argue about the optimal growth directions for the wires.

The method presented in this work is based on the TB models matched not only to reproduce the band structure, but also the correct wavefunction throughout the BZ. Although local approximation to SOC was used here, these models can be straightforwardly extended (to be reported elsewhere) to include non-local SOC effects. The method can be easily extended beyond ZB compounds, and can for example be applied to wurzite InAs nanowires, which are of particular interest in the light of the epitaxially-grown superconductor-semiconductor interfaces~\cite{Krogstrup-NatMat15}.

This method is also optimally suited for the search of other semiconductor materials, suitable for spintronics~\cite{Fabian-RMP04} or Majorana experiments. Further validation of these models versus experimental results suggests the possible route for reliable simulation of realistic devices, similar to those used for realizing exotic topological states which are not accessible to the {\it ab initio} approaches.

\acknowledgements

The authors acknowledge helpful discussions with X. Dai, L. E. Golub, U. Aschauer, Q. S. Wu, G. Winkler, M. Wimmer, P. Krogstrup, M. M. Glazov and E. L. Ivchenko. Special words of gratitude go to M. O. Nestoklon for very instructive suggestions and comments. AS would like to specially thank K. Hummer for useful correspondence. This work was supported by Microsoft Research, the European Research Council through ERC Advanced Grant SIMCOFE, the Swiss National Science Foundation through the National Competence Centers in Research MARVEL and QSIT, and the Aspen Center for Physics through NSF grant PHYD1066293.

\appendix
\section{Parameters for the 14$\times$14 tight-binding model}
\label{sec:tbparameters}
Here we provide the details on the TB model used in this paper. The model is 7$\times$7 (14$\times$14) in the absence (presence) of SOC. It gives a very good description (see Fig.~\ref{fig:InSbbsso}) of the three topmost valence and four lowest conduction bands throughout the BZ.

We describe the way to obtain the model with symmetric parameters using wannier90~\cite{Wannier90} and the hybrid functionals for InSb. However, the same scheme, with little adjustments of the energy window width, that take into account the differences in bandwidths and the fundamental energy gap, works for other binary zincblende semiconductors.
For all materials, experimental lattice constants were used, and the screening parameters of the hybrid functional calculations were adjusted for the best fit of the fundamental gap.

For all the TB models the adopted convention is
\begin{equation}
H_{nm}({\bf k}) = \sum_{\bf R} e^{i{\bf k}\cdot {\bf R}} \langle {\bf 0} n|\hat{H} |{\bf R} m\rangle,
\end{equation}
where the matrix element in the sum stands for the probability amplitude of an electron hopping from orbital $m$ of unit cell ${\bf R}$ to orbital $n$ of the unit cell at the origin {\bf 0}. The summation is limited to local, first- and second-neighbor hoppings. The complete tight binding parameters are available as Supplementary Material.

A hybrid functional calculation is first carried out without SOC. Following the discussion of the Sec.~\ref{sec:tb}~B the outer energy window is fixed to be from $-9$~eV to $10.5$~eV relative to the Fermi level. The inner window is from $-0.3$~eV to $3.1$~eV relative to the Fermi level. The local orbitals for the projection are chosen to be the $s$- and 3$p$- orbitals put on the In site at $(0,0,0)$ and 3$p$-orbitals put on the Sb site at $(\frac{1}{4}, \frac{1}{4}, \frac{1}{4})$ in the units of lattice constant, taken to be $a=6.479${\AA} according to the Ref.~\onlinecite{InSblatt}. No maximal localization is done, only the disentanglement.

As a result one obtains a 7$\times$7 model for InSb without SOC, which is also provided as supplementary material. The resultant WFs are well-localized having the maximum spread $\langle r^2\rangle-\langle r \rangle ^2=4.84$~\AA$^{2}$. Given that the distance to the nearest neighbor is 4.58\AA, this means that to a good approximation the SOC coupling can be considered to be on-site and it also motivates the implementation of external electric field as an on-site energy change.

Local SOC allows to fix time-reversal symmetry by considering a representation to be block block diagonal in spin space. The SOC is generated only by the $p$-orbitals and in the present basis takes the form~\cite{Friedel-JPCS64}
\begin{equation}
H_{\mathrm{SO}}=\frac{\lambda_j}{2}\left(
\begin{array}{cccccc}
0& -i& 0& 0& 0& 1\\
i& 0& 0& 0& 0& -i\\
0& 0& 0& -1& i& 0\\
0& 0& -1& 0& i& 0\\
0& 0& -i& -i& 0& 0\\
1& i& 0& 0& 0& 0\\
\end{array}\right)
\end{equation}
 where $j$ stands for either In or Sb and the matrix is written in $\{p_x\uparrow, p_y\uparrow, p_z\uparrow, p_x\downarrow, p_y\downarrow, p_z\downarrow \}$. The two parameters $\lambda_j$ are then fitted to the {\it ab initio} band structure with SOC included. We used the values $\lambda_{In}=0.226$~eV and $\lambda_{Sb}=0.5181$~eV.

%
%

\bigskip

\section{Scaling of spin splittings with field and slab size}
\label{sec:scaling}

\begin{figure}
\begin{center}
\includegraphics[width=0.9\columnwidth]{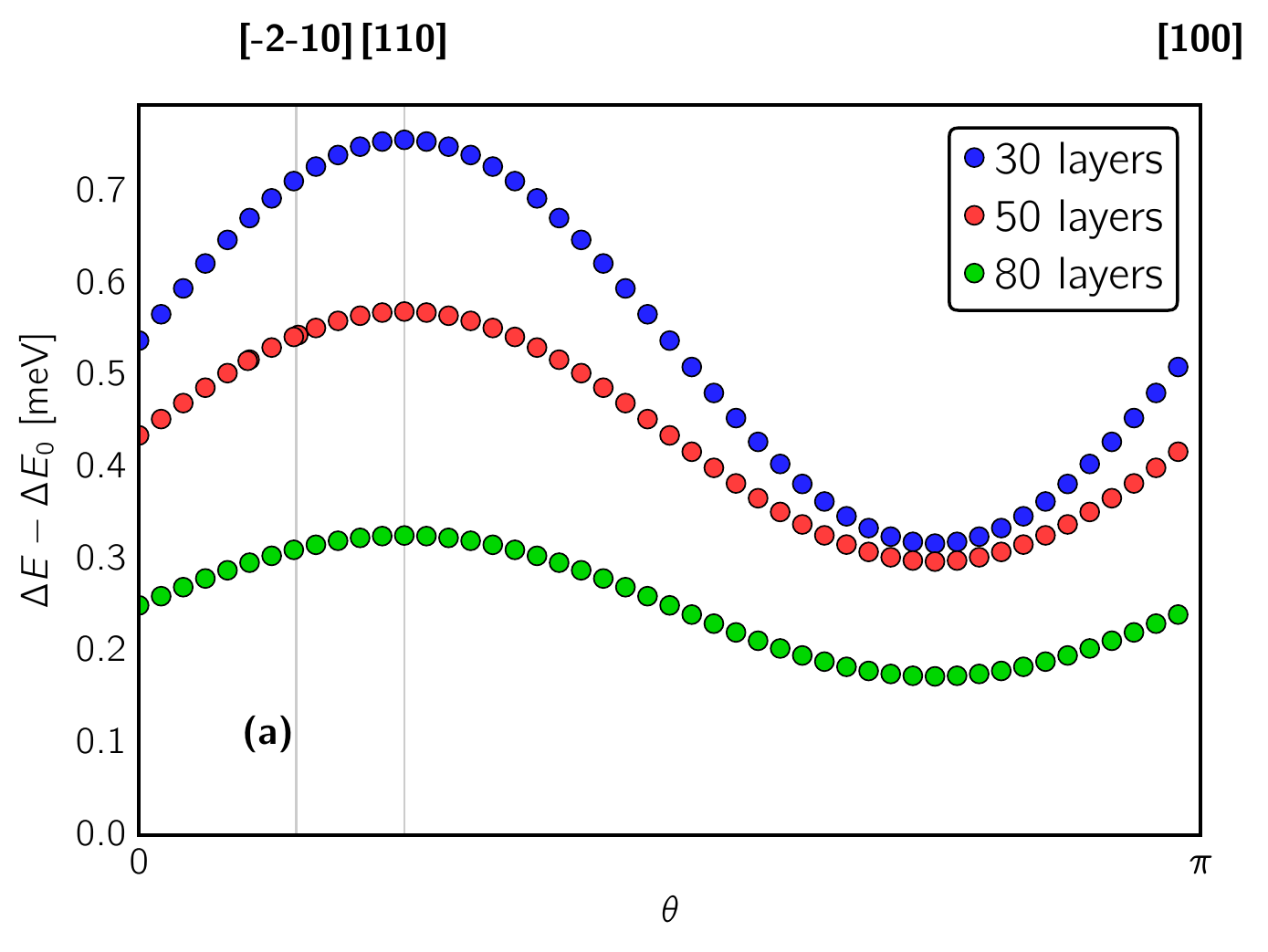}
\includegraphics[width=0.9\columnwidth]{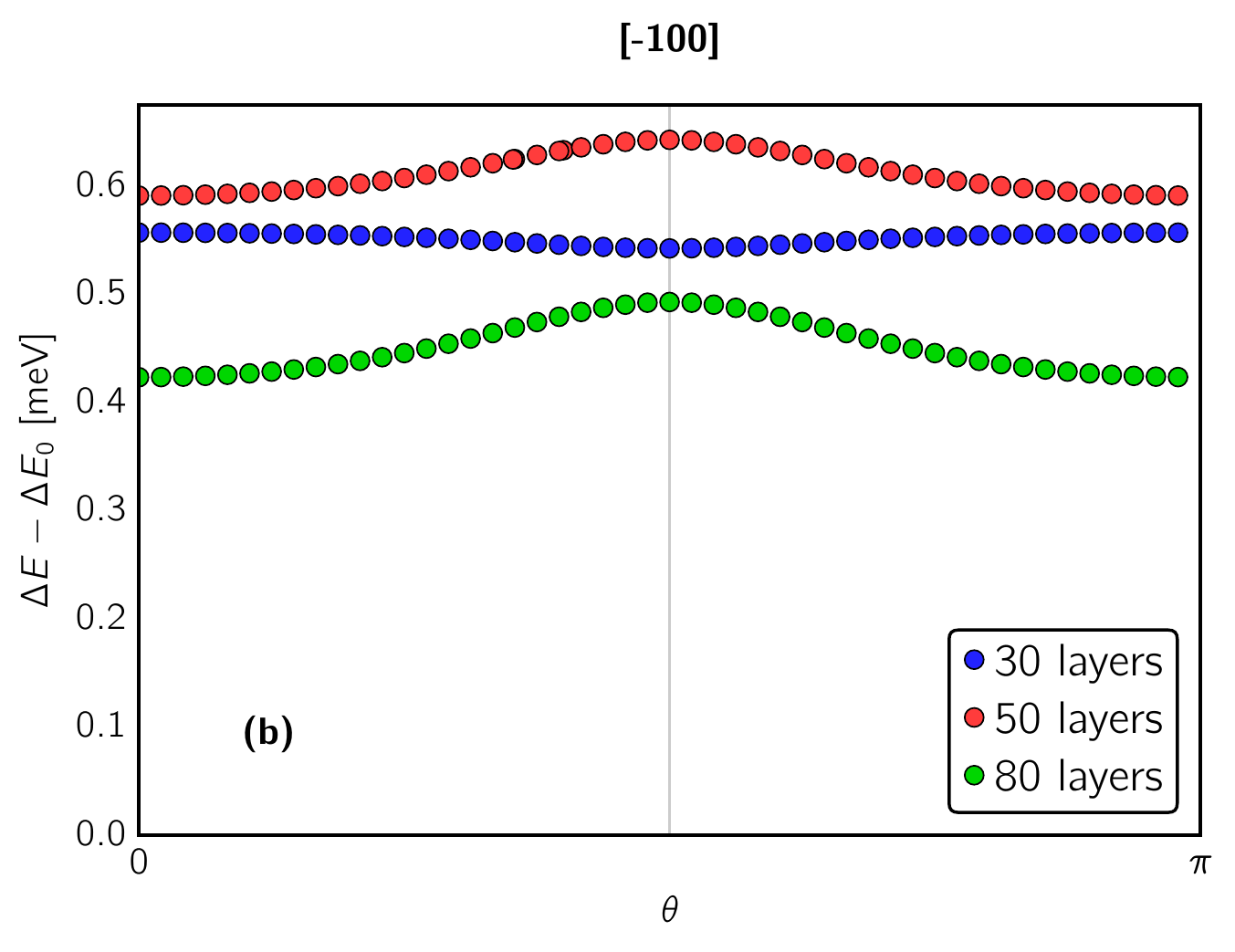}
\includegraphics[width=0.9\columnwidth]{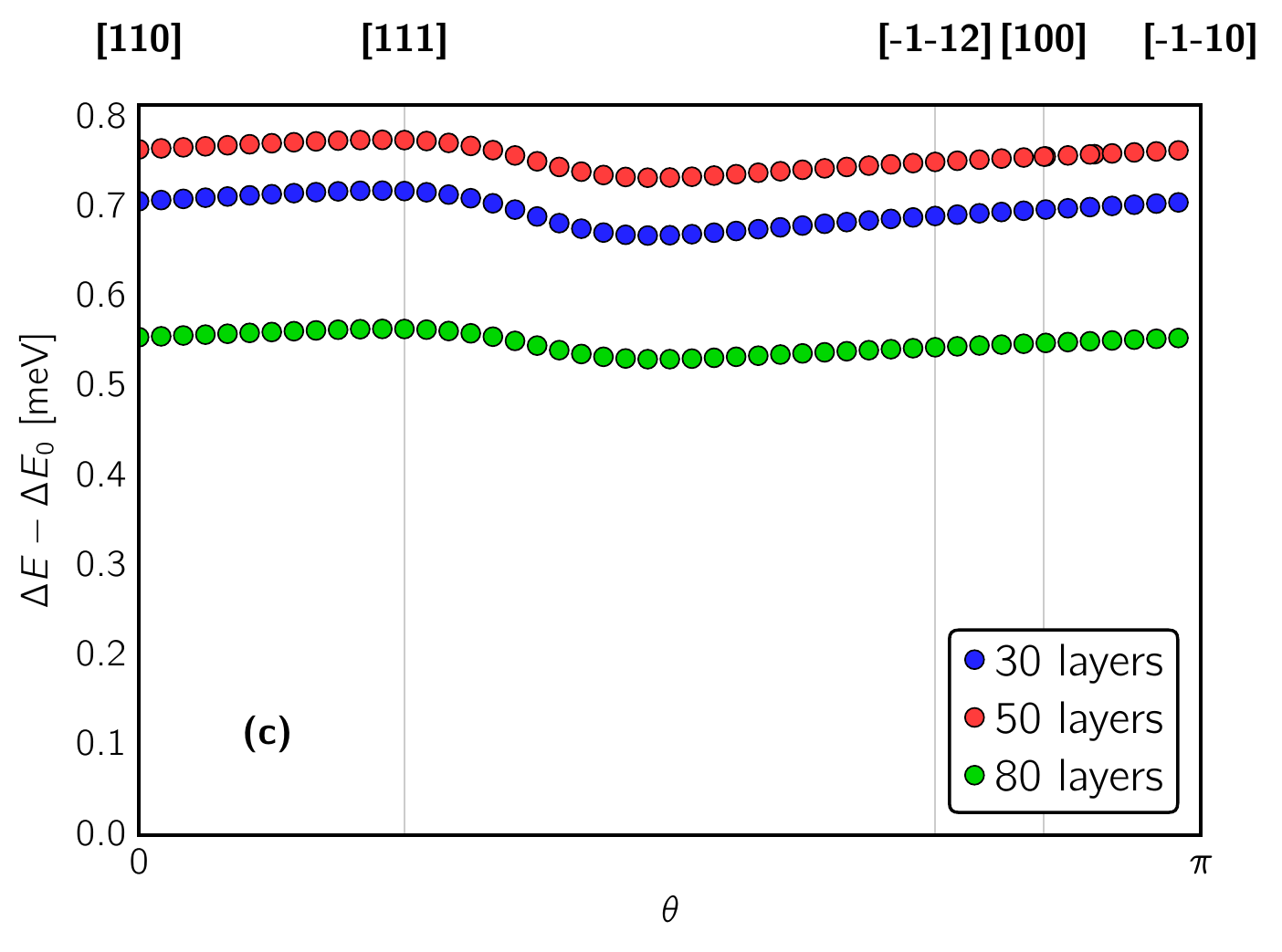}
\end{center}
\caption{Scaling of the field-induced part of the spin splittings $\Delta E_U - \Delta E_0$ with slab thickness. In each case, a potential of $0.4$~V was applied, meaning the field strength was weaker in thicker slabs. The number of unit cells in each slab is 30 (blue) 50 (red) and 80 (green). (a) $(001)$-slab in in the $[001]$ field. (b) $(110)$-slab in in the $[110]$ field. (c) $(1\bar{1}0)$-slab in the $[1\bar{1}0]$-field. }
\label{fig:nscaling}
\end{figure}

\begin{figure}
\begin{center}
\includegraphics[width=0.9\columnwidth]{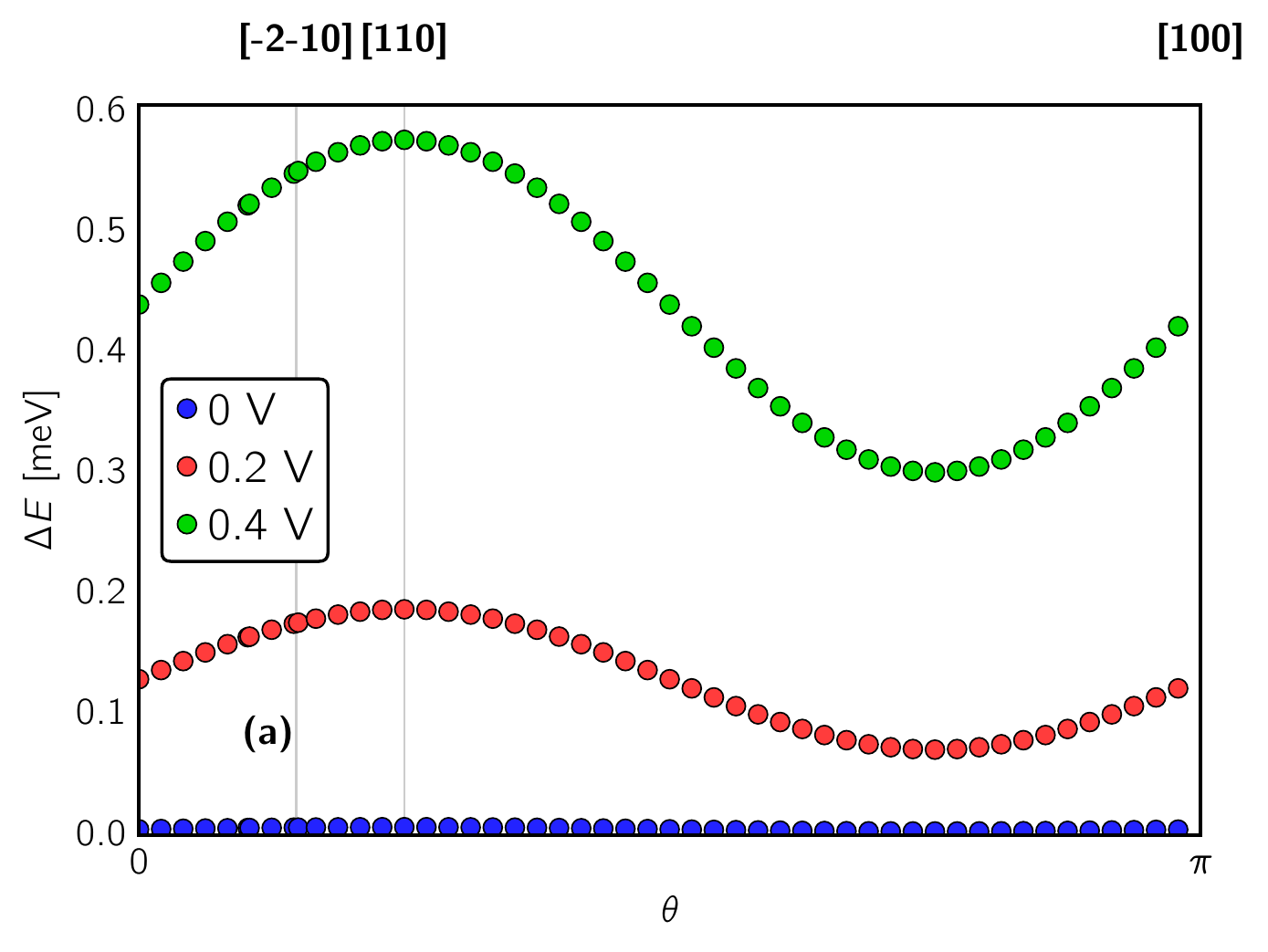}
\includegraphics[width=0.9\columnwidth]{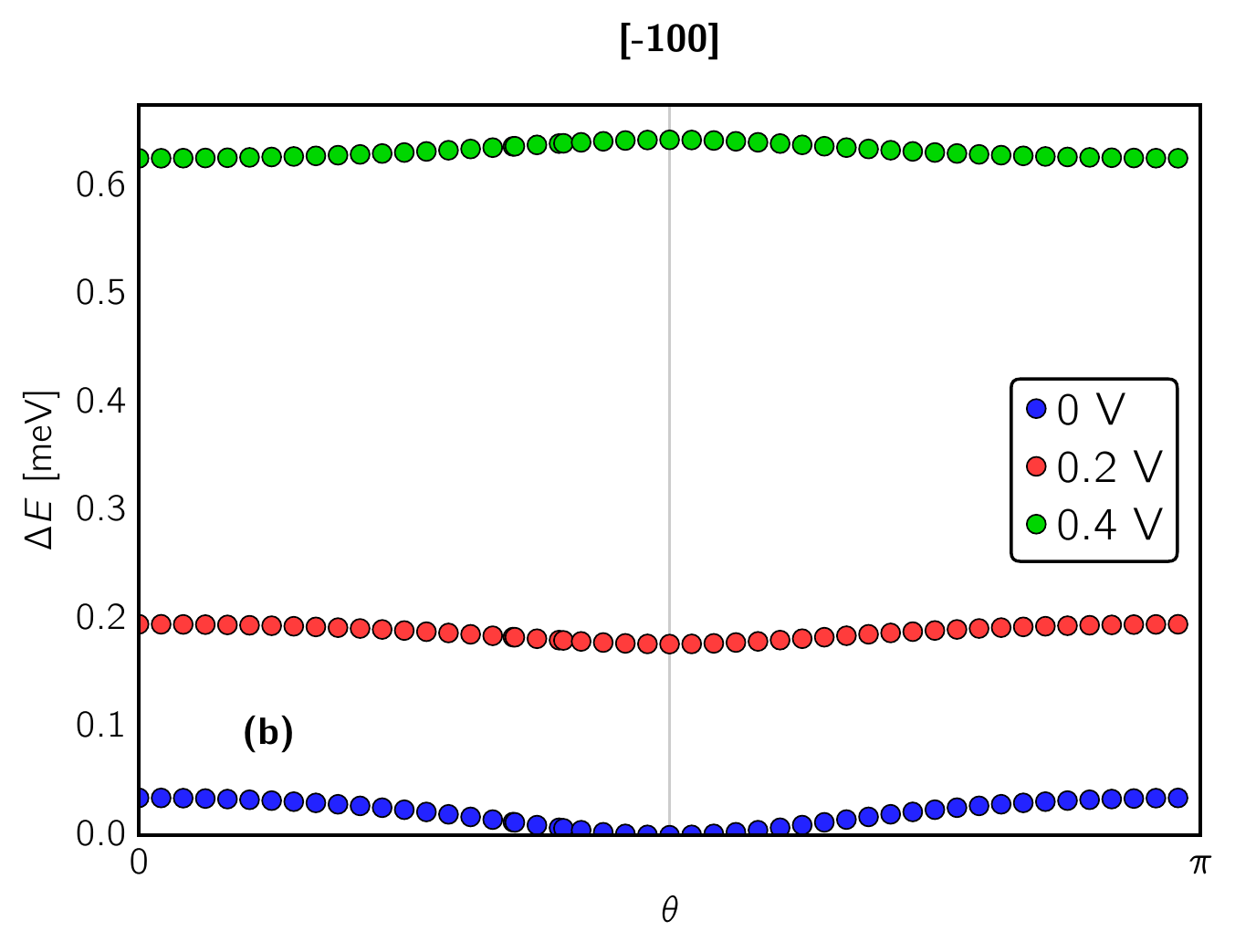}
\includegraphics[width=0.9\columnwidth]{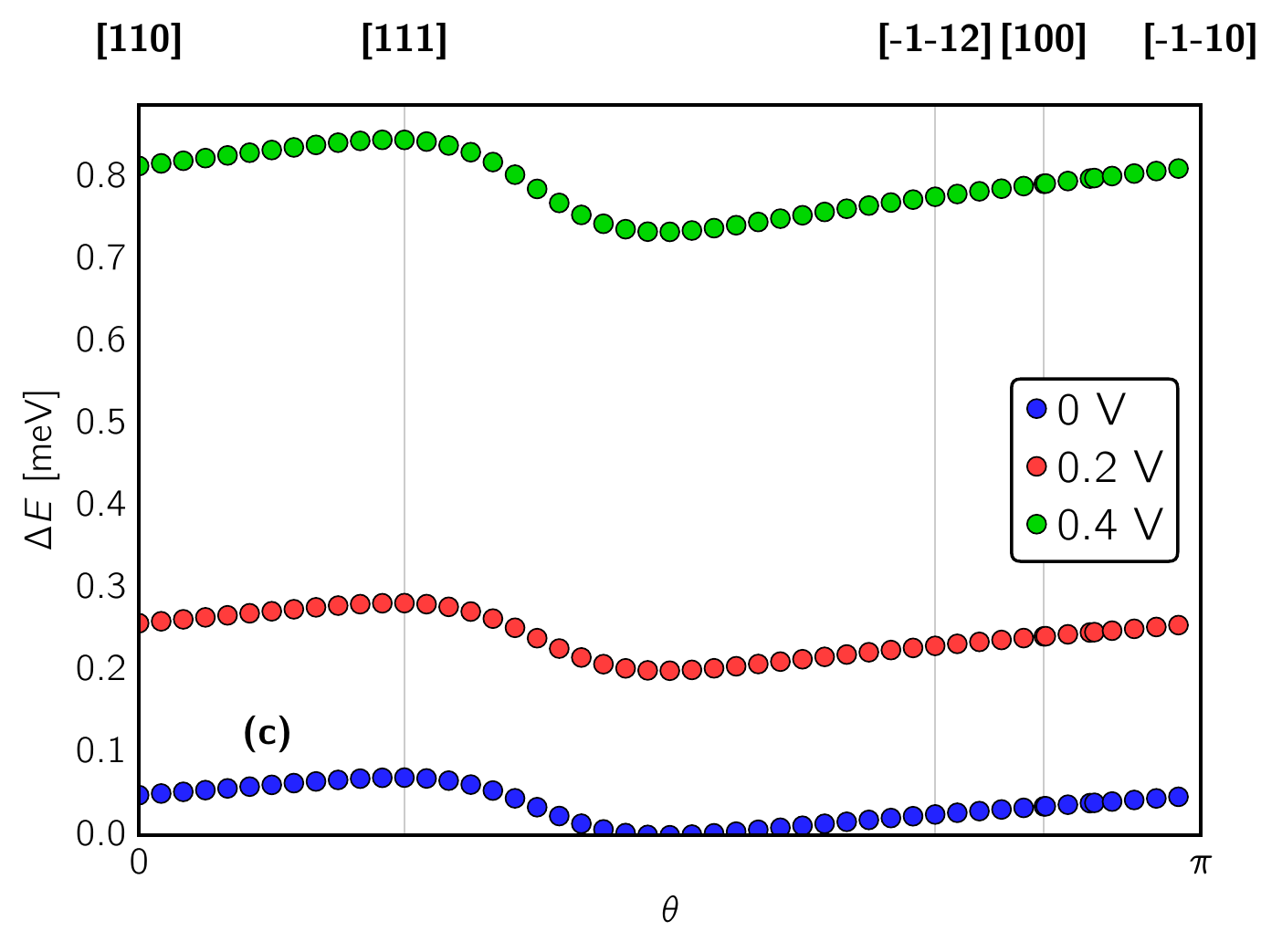}
\end{center}
\caption{Scaling of spin splittings with field. The slab is 50 unit cells wide in all cases. The potential differences across the slab are $0$V (blue), $0.2$V (red) and $0.4$V (green). (a) $(001)$-slab in in the $[001]$ field. (b) $(110)$-slab in in the $[110]$ field. (c) $(1\bar{1}0)$-slab in the $[1\bar{1}0]$-field. }
\label{fig:Uscaling}
\end{figure}
\begin{figure}
\begin{center}
\includegraphics[width=0.9\columnwidth]{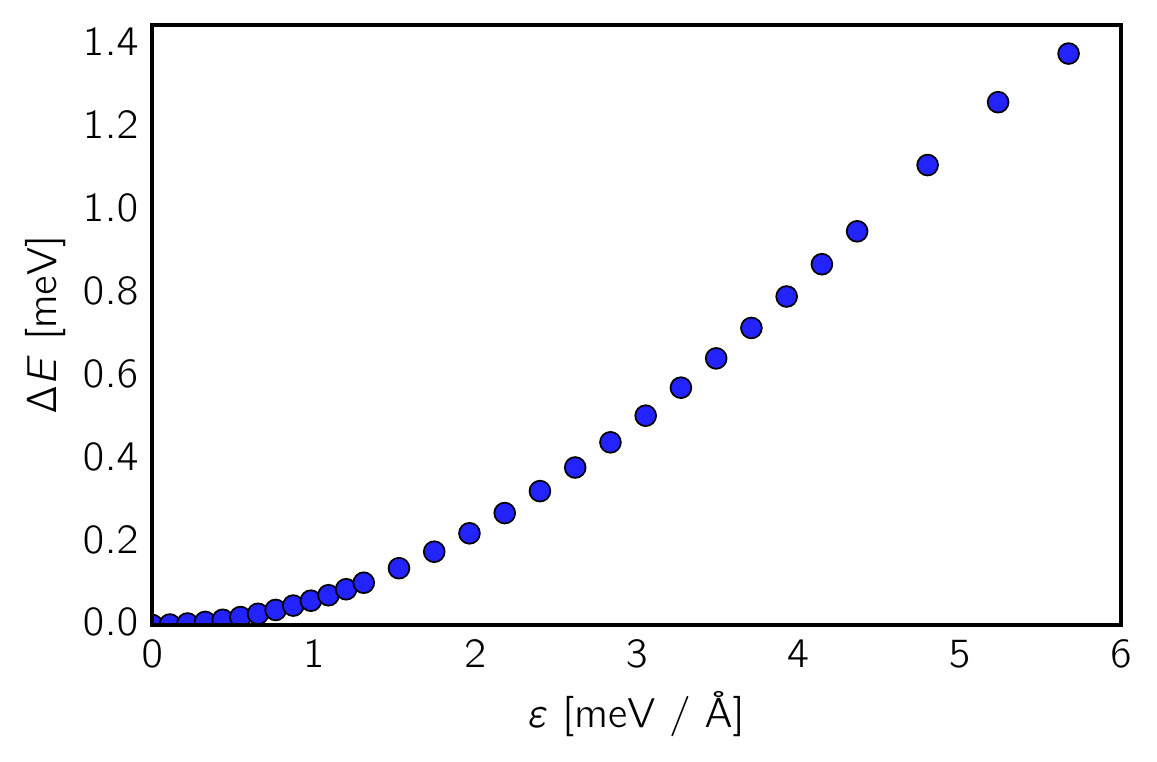}
\end{center}
\caption{Scaling of the total spin splitting $\Delta E$ with field, for a slab of $50$ layers orthogonal to the $[110]$-direction. The splitting is measured in the $[100]$ direction.}
\label{fig:Uscaling_2}
\end{figure}

In this Appendix we present additional data showing how the Rashba SO splitting discussed in Sec.~\ref{sec:so} depends on slab thickness (Fig.~\ref{fig:nscaling}) and strength of the applied field (Fig.~\ref{fig:Uscaling} and Fig.~\ref{fig:Uscaling_2}). It can be seen that the thickness dependence data suggests that the wire thickness should also be taken into account for optimizing spin splittings for realizing MZMs.

\bibliography{zincblende,topological_wires}
\end{document}